\documentclass{jfm}

\usepackage{subfiles}
\usepackage{graphicx}
\usepackage{epstopdf,epsfig,subfloat,subfig,floatrow}
\usepackage{newtxtext}
\usepackage{newtxmath}
\usepackage{natbib}
\usepackage{amsfonts}
\usepackage{amsmath}
\usepackage{amssymb,bm,mathtools}
\usepackage{color}
\usepackage{hyperref}
\usepackage{multirow,makecell,array,booktabs}
\usepackage{soul,cancel}
\usepackage{ulem}
\usepackage{lineno}

\hypersetup{
    colorlinks = true,
    urlcolor   = blue,
    citecolor  = blue,
}

\newcommand{\RomanNumeralCaps}[1]

\allowdisplaybreaks[4]

\definecolor{orange}{RGB}{255,100,0}

\title{Multiscale simulation of coexisting turbulent and rarefied gas flows}

\author{Songyan Tian
        \and Lei Wu
        \corresp{\email{wul@sustech.edu.cn}}
        }

\affiliation{
  Department of Mechanics and Aerospace Engineering, Southern University of Science and Technology, Shenzhen 518055, China
  }

\begin{document}
\maketitle


\begin{abstract}
Simulating complex gas flows from turbulent to rarefied regimes is a long-standing challenge, since turbulence and rarefied flow represent contrasting extremes of computational aerodynamics.
We propose a multiscale method to bridge this gap. 
Our method builds upon the general synthetic iterative scheme for the mesoscopic Boltzmann equation, and integrates the $k$-$\omega$ model in the macroscopic synthetic equation to address turbulent effects. 
We have not only validated this approach using the experimental data of turbulent flows, but also  confirmed that the turbulence model exerts no undue influence on highly rarefied flows. 
Our method is then applied to opposing jet problems in hypersonic flight surrounding by rarefied gas flows, showing that the turbulence could cause significant impacts on the surface heat flux,
which cannot be captured by the turbulent model nor the laminar Boltzmann solution alone. 
This study provides a viable framework for advancing our understanding of the interaction between turbulent and rarefied gas flows.
\end{abstract}

\section{Introduction}

Numerical simulation of aerodynamics has two distinct branches. When the Reynolds number (Re) is large, the Navier-Stokes (NS) equations are used to study turbulence. When the Knudsen number (Kn, the ratio of the molecular mean free path $\lambda$ to the characteristic flow length $L$) is large, the Boltzmann equation is employed to study rarefied flows. The Reynolds number and the Knudsen number are connected through the Mach number (Ma) as
\begin{equation}
\begin{aligned}
 & \text{Re} =\sqrt{\frac{\pi\gamma}{2}} \frac{\text{Ma}}{\text{Kn}}, \quad \text{with} \quad  \text{Ma}=\frac{u}{\sqrt{\gamma RT}}, \quad
  \text{Kn}=\frac{\mu}{pL}\sqrt{\frac{\pi RT}{2}}, \quad
  \text{Re}=\frac{\rho u L}{\mu},
\end{aligned}
\end{equation} 
where $u$ is the characteristic flow velocity; $\gamma$, $R$, $\mu$, $p$, and $\rho$  are the specific heat ratio,  gas constant, shear viscosity, pressure, and density of the gas, respectively. 
Therefore, it is generally believed that the turbulent and rarefied flows do not coexist. 

While the NS equations use the macroscopic quantities (density, flow velocity and temperature) to describe the gas dynamics, the Boltzmann equation employs the velocity distribution function (VDF). 
From the physical perspective, the Boltzmann equation is more elemental, as according to the Chapman-Enskog expansion~\citep{chapman1990mathematical}, the NS equations can be derived from the Boltzmann equation in the continuum flow regime, where Kn is small. When Kn increases, the gas flow gradually falls into the transition and free-molecular flow regimes, where the traditional laws of viscosity and heat conduction in the NS equations become inapplicable, necessitating the use of the Boltzmann equation. However, from the numerical perspective, since the VDF is defined in the six-dimensional phase-space, 
the Boltzmann equation, which is solved by the stochastic direct simulation Monte Carlo (DSMC) method~\citep{bird1994molecular} or deterministic discrete velocity methods~\citep{Aristov2001}, is typically applied in laminar rarefied gas flows, where the number of spatial grids is much less than that used in the direct numerical simulation (DNS) of NS equations. 

There are many length scales in a turbulent flow. If $L$ is chosen to be the smallest Kolmogorov length scale, the local Knudsen number can be large, and the NS equations may fail to describe the full multiscale turbulence. Many researchers have been trying to find the rarefaction effects in turbulent flows~\citep{CERCIGNANCI1992,CERCIGNANCI1993}. For instances, \cite{Komatsu2014} conducted the large scale molecular dynamics simulations of freely decaying turbulence, but ``the energy spectrum is observed to scale reasonably well according to Kolmogorov scaling, even though the Kolmogorov length is of the order of the molecular scale". \cite{Gallis2017} used the DSMC method to simulate the Taylor-Green vortex flow, and found it ``reproduces the Kolmogorov law and agree well with the turbulent kinetic energy and energy dissipation rate obtained from DNS of the NS equations''.
\cite{Li2018} simulated the decaying isotropic turbulence by solving the Boltzmann kinetic equation via the deterministic scheme of~\cite{Xu2010}, and found that the kinetic equation predicts the energy-decaying exponent about 10\% larger than the NS equations.
However, the difference in statistical quantities is only a few percent~\citep{Li2018PhD}.  
\cite{Gallis2022PRL} discovered that the sole divergence in the energy spectrum from the DSMC and NS simulation results occurs at large wave numbers, where thermal fluctuations induce the energy spectrum to scale proportionally to the square of wave number\footnote{This is also observed in the molecular dynamics simulation by~\cite{Komatsu2014}, and can be predicted by the fluctuating NS equations~\citep{FNS}. Therefore, these results cannot be attributed to the rarefaction effects in turbulence.}. 
They also reported that ``thermal fluctuations have little impact on the large-scale evolution of the flow''.

To date, no instances of coexisting turbulence and rarefied flows with ``engineering significance" have been identified. That is, in terms of numerical simulations, there have been no instances where the stress and heat flux have significantly deviated from the values predicted by the NS equations. The absence of such cases might be attributed to the time-intensive nature of DNS based on the Boltzmann equation and even the more time-consuming molecular dynamics simulation, which limits the range of flow scenarios and parameters explored.
Therefore, this paper is dedicated to developing a multiscale method to simulate the gas flow spanning from turbulent to rarefied conditions, and to identifying engineering scenarios 
where turbulence and rarefaction coexist.

\section{A multiscale model from the turbulent to rarefied flows }\label{sec:NumMethod}

Although the Boltzmann equation encapsulates the gas dynamics  across turbulent to rarefied regimes, the computational demands of DNS are overwhelming. In regions of high Re, the spatial grid must be finely resolved, while in regions of high Knudsen number Kn, the discrete velocity grid requires dense sampling. Given that the DNS of NS equations is already a time-intensive process for simulating turbulence, the prospect of performing the DNS of the Boltzmann equation in scenarios where turbulence and rarefaction coexist is beyond the reach of our computational capabilities. 
To address this challenge, we introduce a multiscale modeling approach that offers a viable alternative.
Our idea is to couple the Reynolds-averaged Navier–Stokes (RANS) turbulence model into the Boltzmann equation, to strike a balance between the accuracy and efficiency. The coupling should process the asymptotic-preserving property, i.e., it automatically recovers the RANS model when Re is high, and the Boltzmann equation when Kn is large. Moreover, from the numerical perspective, since the RANS produces the time-averaged steady state, the Boltzmann equation will also be solved by the general synthetic iteration scheme (GSIS) tailored for efficient steady-state simulations~\citep{Su2020,Zhang2024}.

In the following, we first introduce the Boltzmann kinetic model which describe the gas dynamics from the continuum to rarefied flow regimes. Second, we introduce the two-temperature NS equations derive from the kinetic equation when the Knudsen number is small, and describe the RANS model for efficient simulation of turbulence on coarse spatial grid. 
Third, we couple the GSIS and RANS to efficiently find the solution of the kinetic equation from the turbulent to rarefied  regimes.



\subsection{The gas kinetic model}

While the dynamics of monatomic gases is described by the Boltzmann equation, that of the polyatomic gas is described by the \cite{wangcs1951transport} equation. 
To reduce the computational cost, simplified kinetic equations are usually adopted in numerical simulations \citep{Wu2015,li2021uncertainty}. Here, two reduced VDFs $f_0(t, \bm{x}, \bm{\xi})$ and $f_1(t, \bm{x}, \bm{\xi})$ are used to describe the translational and rotational states of molecular gas, where $t$ is the time, $\bm{x}=(x_1,x_2,x_3)$ is the spatial coordinate, $\bm{\xi}=(\xi_1,\xi_2,\xi_3)$ is the molecular velocity. The rotational degrees-of-freedom is a constant $d_r$, and it is assumed the vibrational and electronic energy levels are not excited, although the model can be easily extended to include these non-equilibrium and even multi-physics effects such as radiation~\citep{LiQi2023JFM}. Macroscopic quantities, such as the density $\rho$, flow velocity $\bm{u}$, deviatoric stress $\bm{\sigma}$, translational and rotational temperature $T_t$ and $T_r$, translational and rotational heat flux $\bm{q}_t$ and $\bm{q}_r$, are obtained by taking the moments of VDFs:
\begin{equation}\label{eq:getmoment}
    \begin{aligned}
  \left(\rho,~\rho\bm{u},~\bm{\sigma},~\frac{3}{2}\rho RT_t,~\bm{q}_{t}\right) & =\int\left(1,~\bm{\xi},~\bm{c}\bm{c}-\frac{c^2}{3}\mathrm{I},~\frac{c^2}{2},~\frac{c^2}{2}\bm{c}
        \right) f_0 \mathrm{d}\bm{\xi},\\
 \left(\frac{d_r}{2}\rho RT_r,~\bm{q}_{r}\right) & =\int\left(1,~\bm{c}\right)f_1\mathrm{d}\bm{\xi},
    \end{aligned}
\end{equation}
where $\bm{c}=\bm{\xi}-\bm{u}$ is the peculiar (thermal) velocity, and $\textbf{I}$ is the identity matrix. The pressure related to the translational motion is $p_t=\rho R T_t$, while the total pressure is $p=\rho RT$.

The evolution of VDFs is governed by the following Boltzmann kinetic equations:
\begin{equation}\label{general_model}
    \begin{aligned}
        &\frac{\partial f_0}{\partial t}+\bm{\xi}\cdot \nabla f_0 = \frac{g_{0t}-f_0}{\tau}+\frac{g_{0r}-g_{0t}}{Z_r\tau}, \\
        & \frac{\partial f_1}{\partial t}+\bm{\xi}\cdot \nabla f_1 = \frac{g_{1t}-f_1}{\tau}+\frac{g_{1r}-g_{1t}}{Z_r\tau}, 
    \end{aligned}
\end{equation}
where $\tau$ and ${Z}_{r}\tau$ are the elastic and inelastic collision characteristic time, respectively, with ${Z}_{r}$ being the rotational collision number. The elastic collision conserves the kinetic energy, while the inelastic collision exchanges the translational and rotational energies. In order the make the kinetic model mimic the behaviors (such as to recover the shear viscosity, bulk viscosity, translational and internal heat conductivities) of the \cite{wangcs1951transport} equation as closely as possible, the reference VDFs to which the gas system will relax are designed as follows:
\begin{equation}\label{reference_VDF}
    \begin{aligned}
        g_{0t}&= \rho\left(\frac{1}{2\pi RT_t}\right)^{3/2}\exp\left(-\frac{c^2}{2RT_t}\right)\left[1+\frac{2\bm{q}_{t}\cdot\bm{c}}{15RT_tp_t}\left(\frac{c^2}{2RT_t}-\frac{5}{2}\right)\right],\\
        g_{0r}&= \rho\left(\frac{1}{2\pi RT}\right)^{3/2}\exp\left(-\frac{c^2}{2RT}\right)\left[1+\frac{2\bm{q}_{0}\cdot\bm{c}}{15RTp}\left(\frac{c^2}{2RT}-\frac{5}{2}\right)\right],\\
        g_{1t}&=\frac{d_r}{2}RT_rg_{0t} + \left(\frac{1}{2\pi RT_t}\right)^{3/2}\frac{\bm{q}_{r}\cdot\bm{c}}{RT_t}\exp\left(-\frac{c^2}{2RT_t}\right), \\
        g_{1r}&=\frac{d_r}{2}RTg_{0r} + \left(\frac{1}{2\pi RT}\right)^{3/2}\frac{\bm{q}_{1}\cdot\bm{c}}{RT}\exp\left(-\frac{c^2}{2RT}\right),
    \end{aligned}
\end{equation}
with $\bm{q}_{0},~\bm{q}_{1}$ being linear combinations of translational and internal heat fluxes \citep{li2021uncertainty}:
\begin{equation}
    \begin{bmatrix} 
        \bm{q}_{0} \\ \bm{q}_{1} 
    \end{bmatrix}
    =
    \begin{bmatrix}		
        (2-3A_{tt})Z_r+1 & -3A_{tr}Z_r  \\		
        -A_{rt}Z_r & -A_{rr}Z_r+1 \\ 
    \end{bmatrix}
    \begin{bmatrix} 
    \bm{q}_{t} \\ \bm{q}_{r} 
    \end{bmatrix},
\end{equation}
where $\bm{A}=[A_{tt},A_{tr},A_{rt},A_{rr}]$ is determined by the relaxation rates of heat flux. For nitrogen, we choose $Z=3.5$ and $\bm{A}=[0.786,-0.201,-0.059,0.842]$.

The gas kinetic model describes the gas-gas interaction, in order to fully determine the gas dynamics in wall-bounded problems, the gas-surface interaction should be specified. In this work, we adopt the Maxwell gas-surface boundary condition, where the gas molecules hitting the solid wall will be reflected diffusely.

\subsection{The NS equations and the RANS model}\label{sec:TurbModel}

Exploiting the relation between macroscopic quantities and mesoscopic VDFs in \eqref{eq:getmoment}, the macroscopic moment equations can be derived from the kinetic equation \eqref{general_model} as follows:
\begin{equation}\label{eq:macroscopic_equation_2}
\begin{aligned}
\frac{\partial \rho}{\partial t} + \nabla \cdot (\rho \boldsymbol{u}) = 0,\\
\frac{\partial \rho \boldsymbol{u}}{\partial t} + \nabla \cdot (\rho \boldsymbol{u} \boldsymbol{u}) = -\nabla \cdot (\rho R T_t \textbf{I} + \boldsymbol{\sigma}),\\    
\frac{\partial \rho e}{\partial t} + \nabla \cdot (\rho e \boldsymbol{u}) = -\nabla \cdot (\rho R T_t \boldsymbol{u} + \boldsymbol{\sigma} \cdot \boldsymbol{u} + {\boldsymbol{q}}_{t} + {\boldsymbol{q}}_{r}),\\
\frac{\partial \rho e_r}{\partial t} + \nabla \cdot (\rho e_r \boldsymbol{u}) = -\nabla \cdot {\boldsymbol{q}}_{r} + \frac{d_r \rho R}{2}\frac{T-T_r}{{Z}_{r}\tau},
\end{aligned}
\end{equation}
where the first three equations respectively describe the conservation of mass, momentum, and total energy, while the fourth equation accounts for the exchange between translational and rotational energies. 

Clearly, \eqref{eq:macroscopic_equation_2} is not closed, since expressions for the stress and heat fluxes have not been expressed in terms of the density, velocity, temperature and their gradients. The Chapman-Enskog expansion of the kinetic equation to the first-order of Kn (usually valid when Kn $\lessapprox0.01$) yields the NS constitutive relations as:
\begin{equation}\label{eq:NSF_constitutive}
    \begin{aligned}
        \bm{\sigma}^{\text{NS}} &= -\mu \left(\nabla\bm{u}+\nabla\bm{u}^{\mathrm{T}}-\frac{2}{3}\nabla\cdot\bm{u}\mathrm{I}\right),\\
        \bm{q}_{t}^{\text{NS}} &= -\kappa_t\nabla T_t,\quad
        \bm{q}_{r}^{\text{NS}} = -\kappa_r\nabla T_r,
    \end{aligned}
\end{equation}
where the (laminar, or physical) shear viscosity is obtained as $\mu=p_t\tau$, while the transitional and rotational thermal conductivities $\kappa_t$ and $\kappa_r$ are~\citep{li2021uncertainty}:
\begin{equation}\label{eq:kappaAnumber}
	\left[ 
      \begin{array}{ccc} 
        \kappa_t \\ \kappa_r 
      \end{array}
    \right]
	= \frac{\mu}{2}
	\left[ 
      \begin{array}{ccc} 
        A_{tt} & A_{tr} \\ A_{rt} & A_{rr} 
      \end{array}
    \right]^{-1}
    \left[ 
      \begin{array}{ccc} 
        5 \\ d_r 
      \end{array}
    \right]
        = \mu
        \left[
        \begin{array}{ccc} 
        3.546 \\ 1.435
        \end{array}
        \right].
\end{equation}

Theoretically, turbulence can be investigated by the DNS of \eqref{eq:macroscopic_equation_2} with the NS constitutive relations~\eqref{eq:NSF_constitutive}. However, its prohibitive numerical cost prompts the use of turbulence models. 
Though the RANS models are derived on coarse grids under multiple assumptions, they are widely accepted in engineering due to its low computational cost and good accuracy on capturing the flow pattern. 
Here, the popular $k$-$\omega$ shear stress transport (SST) model, which combines the advantages of both the low- and high-Reynolds numbers $k$-$\omega$ models and embeds the Bradshaw's assumption for boundary layer, is considered~\citep{Menter1994}. It is good at predicting of flow with adverse pressure gradients and separation.

In the SST model, the moment equations \eqref{eq:macroscopic_equation_2} are used, but the constitutive relations not only contains the physical viscosity (heat conductivity), but also the turbulent ones. That is, $\mu$ and $\kappa$ (irrespective of translational or rotational) in \eqref{eq:NSF_constitutive} are modified as 
\begin{equation}\label{eq:mutCoupling}
\begin{aligned}
    & \mu = \mu_{lam} + \mu_{turb}, \\
    & \kappa = \kappa_{lam} + \kappa_{turb} = \frac{\mu_{lam} c_p}{\text{Pr}_{lam}} +  \frac{\mu_{turb} c_p}{\text{Pr}_{turb}} =  \left(\mu_{lam} + \frac{\mu_{turb}\text{Pr}_{lam}}{\text{Pr}_{turb}}\right) \frac{c_p}{\text{Pr}_{lam}},
\end{aligned}
\end{equation}
where the subscript $turb$ denotes the turbulent parts, and the laminar to turbulent Prandtl number ratio is set to 0.8, with the value of $c_p/\text{Pr}_{lam}$ for $\kappa_t$ and $\kappa_r$ obtained with \eqref{eq:kappaAnumber} accordingly.

The dimensionless turbulent viscosity in \eqref{eq:mutCoupling} is given by\footnote{$a_R$ is a constant from the Bradshaw assumption, $F_\mu$ is a blending function that takes the value of one for boundary layer flows and zero for free shear flows. $F_\phi$ is the blending function which transform the model and its coefficients from $k$-$\omega$ model in near wall region ($F_\phi=1$) to $k$-$\epsilon$ model in the main flow area ($F_\phi=0$), the model coefficient $\phi=(\sigma_k, \sigma_{\omega}, \beta, \beta^*, \kappa_{\phi}, \gamma)$ is blended as $\phi=F_\phi \phi_1 + (1-F_\phi) \phi_2$. These model functions and coefficients are given in \cite{Menter1994}. }
\begin{equation}\label{eq:mu_t}
\mu_{turb} ={a_R \rho k}/{\max\left(a_R \omega, \frac{\Vert\Omega\Vert F_{\mu}}{\text{Re}_{ref}}\right)},
\end{equation}
where $k$ represents the turbulent kinetic energy, $\omega=\rho k/\mu_t$ is the turbulent dissipation frequency, with $\Vert\Omega\Vert$ being the vorticity magnitude.
They can be calculated by the following equations\footnote{
To be consistent with the normalization in GSIS~\citep{Zhang2024}, reference quantities are defined based on free flow properties (denoted by subscript 0): density $\rho_{ref}=\rho_0$, temperature $T_{ref}=T_0$, velocity $v_{ref}=\sqrt{RT_0}$, pressure $p_{ref}=\rho_0 R T_0$, heat flux $q_{ref}=p_{ref}v_{ref}$, viscosity $\mu_{ref}=\rho_{ref}v_{ref}L$, turbulent kinetic energy $k_{ref}=v^2_{ref}$, turbulent dissipation frequency $\omega_{ref}=\rho_{ref}v^2_{ref}/\mu_{ref}$.
}:
\begin{equation}\label{eq:k}
\frac{\partial{\rho k}}{\partial{t}}+\frac{\partial{\rho u_j k}}{x_j}=\frac{1}{\text{Re}_{ref}} \frac{\partial}{\partial{x_j}}
\left[
(\mu_{lam}+\sigma_k \mu_{turb})\frac{\partial{k}}{\partial{x_j}}
\right]
+\text{Prod}
-\underbrace{\text{Re}_{ref}\beta^* \rho \omega k}_{Diss_k},
\end{equation}
\begin{equation}\label{eq:omega}
\begin{aligned}
\frac{\partial{\rho \omega}}{\partial{t}}+\frac{\partial{\rho u_j \omega}}{x_j}=&\frac{1}{\text{Re}_{ref}} \frac{\partial}{\partial{x_j}}
\left[
(\mu_{lam}+\sigma_\omega \mu_{turb})\frac{\partial{\omega}}{\partial{x_j}}
\right]
+\gamma \frac{\rho}{\mu_{turb}} \tau_{turb,ij}\frac{\partial{u_i}}{\partial{x_j}}\\
&-\text{Re}_{ref}\beta \rho \omega^2
+\frac{2 (1-F_\phi)}{\text{Re}_{ref}} \frac{\rho \sigma_{\omega 2}}{\omega} \frac{\partial{k}}{\partial{x_j}}\frac{\partial{\omega}}{\partial{x_j}}.
\end{aligned}
\end{equation}
Note that $\boldsymbol{\tau}_{turb} = \mu_{turb} \left(\nabla{\boldsymbol{u}}+\nabla{\boldsymbol{u^\text{T}}}-\frac{2}{3}\nabla\cdot\boldsymbol{u}\textbf{I}\right) - \frac{2}{3} \rho k \textbf{I}$ is the Favre-averaged Reynolds-stress computed under the Boussinesq eddy-viscosity hypothesis. To eliminate certain erroneous spikes of $\mu_{turb}$ rooted in two-equation turbulence models~\citep{Menter1993}, the production term in \eqref{eq:k} is 
$\text{Prod}= \min \left(\tau_{turb,ij}\frac{\partial{u_i}}{\partial{x_j}}, 20 Diss_k\right)$.
At solid surface, $k=0$ and $\omega =({60\mu_{lam}})/({\text{Re}_{ref}{\beta}_{1}\rho D_{1}^{2}})$ are specified, with $D_{1}$ being the distance of the first cell center to the wall.

The SST model, originally developed for incompressible flows, does not account for density fluctuations. However, in hypersonic or high heat transfer scenarios, three terms in the density-weighted Favre-averaged equation for turbulent kinetic energy are not modeled: pressure-dilatation, pressure work, and curl-free dilatation dissipation. These terms are crucial as they contribute to the reduction of turbulent kinetic energy in flows with $\text{Ma}>5$. 
In this study, the pressure-dilatation and dilatation dissipation terms are taken into account as per \cite{Sarkar1992}, while the pressure work is discarded due to the lack of satisfactory correction model and proof of correctness for coupling differently established correction models of different terms. 
The final form of the implemented SST model reads:
\begin{equation}
\begin{aligned}
\frac{\partial{\rho k}}{\partial{t}}+\frac{\partial{\rho u_j k}}{x_j}=&\frac{1}{\text{Re}_{ref}} \frac{\partial}{\partial{x_j}}
\left[
(\mu_{lam}+\sigma_k \mu_{turb})\frac{\partial{k}}{\partial{x_j}}
\right]
+\text{Prod}-\underbrace{\text{Re}_{ref}\beta^* \rho \omega k}_{Diss_k}\\
&-\underbrace{\text{Re}_{ref}\beta^*  \xi^* M^2_t \rho \omega k}_{\text{Dilatation dissipation}}
+\underbrace{\alpha_2 \tau_{turb,ij}\frac{\partial{u_i}}{\partial{x_j}} M_t + \text{Re}_{ref} \alpha_3 \beta^* \rho \omega k M^2_t}_{\text{pressure dilatation}},
\end{aligned}
\end{equation}
\begin{equation}\label{eq:omega}
\begin{aligned}
\frac{\partial{\rho \omega}}{\partial{t}}+\frac{\partial{\rho u_j \omega}}{x_j}=&\frac{1}{\text{Re}_{ref}} \frac{\partial}{\partial{x_j}}
\left[
(\mu_{lam}+\sigma_\omega \mu_{turb})\frac{\partial{\omega}}{\partial{x_j}}
\right]
+\gamma \frac{\rho}{\mu_{turb}} \tau_{turb,ij}\frac{\partial{u_i}}{\partial{x_j}}\\
&-\text{Re}_{ref}\beta \rho \omega^2
+\frac{2 (1-F_\phi)}{\text{Re}_{ref}} \frac{\rho \sigma_{\omega 2}}{\omega} \frac{\partial{k}}{\partial{x_j}}\frac{\partial{\omega}}{\partial{x_j}}
+\underbrace{\text{Re}_{ref}\beta^* \xi^* M^2_t \rho \omega^2}_{\text{Dilatation dissipation}},
\end{aligned}
\end{equation}
where $M_t=\sqrt{2k}/\sqrt{\gamma R T}$ is the turbulent Mach number. Such corrections work fine for jets and free shear flow at low Reynolds number. 
Other model coefficients ($\alpha_2, \alpha_3, \xi^*$) are summarized by~\cite{Wilcox2006turbulence}.



\subsection{ Coupling the GSIS with RANS model}

To investigate the rarefied gas dynamics, the kinetic equation~\eqref{general_model} should be solved numerically. However, since the VDFs are defined in six-dimensional phase space, the computational cost is much higher than the NS equations. Fortunately, since in most engineering problems the rarefied gas flows are laminar, only the steady states are considered. Therefore, the implicit iteration methods are normally used. Here we briefly introduce the recently developed GSIS for efficiently simulating the rarefied laminar flows. 

GSIS is a multi-scale method that alternatively solves the mesoscopic kinetic equation and macroscopic synthetic equation. The synthetic equation is given by \eqref{eq:macroscopic_equation_2}, but now the stress and heat flux are constructed as
\begin{equation}\label{eq:full_constitutive}
    \begin{aligned}
        \bm{\sigma} &= -\mu \left(\nabla\bm{u}+\nabla\bm{u}^{\mathrm{T}}-\frac{2}{3}\nabla\cdot\bm{u}\mathrm{I}\right) + \text{HoT}_{\bm{\sigma}},\\
        \bm{q}_{t} &= -\kappa_t\nabla T_t + \text{HoT}_{\bm{q}_{t}},\quad
        \bm{q}_{r}= -\kappa_r\nabla T_r + \text{HoT}_{\bm{q}_{r}},
    \end{aligned}
\end{equation}
where the high-order terms (HoTs) describing the rarefaction effects are constructed as:
\begin{equation}\label{eq:getHoTs}
    \begin{aligned}
        \text{HoT}_{\bm{\sigma}} &= \int \left(\bm{c}\bm{c}-\frac{c^2}{3}\mathrm{I}\right)f_0 \mathrm{d}\bm{\xi} -\bm{\sigma}^{\text{NS}},\\
      \text{HoT}_{\bm{q}_{t}} &= \int \bm{c} \frac{c^2}{2}f_0 \mathrm{d}\bm{\xi} -\bm{q}_{t}^{\text{NS}},\quad
        \text{HoT}_{\bm{q}_{r}} = \int \bm{c}f_1 \mathrm{d}\bm{\xi}  -\bm{q}_{r}^{\text{NS}},
    \end{aligned}
\end{equation}
with $\bm{\sigma}^{\text{NS}}$, $\bm{q}_{t}^{\text{NS}}$ and $\bm{q}_{r}^{\text{NS}}$ calculated based on \eqref{eq:NSF_constitutive} using the macroscopic properties from the moments of the VDFs as well. 

\begin{figure}[t]
\centering
\includegraphics[trim={0 20 0 10},clip,width=0.7\textwidth]{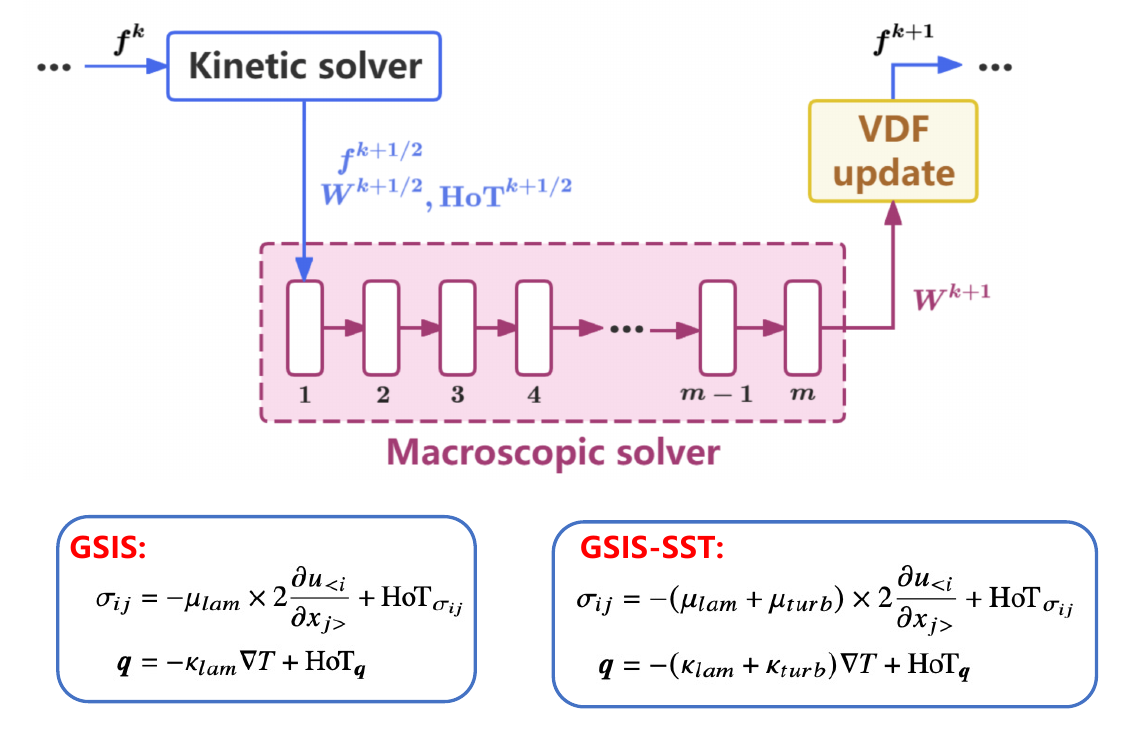}
\caption{Schematic of the GSIS and GSIS-SST. The kinetic equation is solved by the discrete velocity method iteratively (here $k$ is the iteration step). Between each iteration, the HoTs extracted from the kinetic solver are fed into the macroscopic synthetic equation, which is solved to the steady state or for $m\approx200$ steps. Then, the updated macroscopic quantities are fed into the kinetic solver through the modification of VDFs. Detailed implementation of GSIS are given by~\cite{Liu2024}. Different to the pure GSIS, in GSIS-SST the turbulent viscosity and heat conductivities are added in the macroscopic synthetic equation.
}
\label{Demo}
\end{figure}

As sketched in figure~\ref{Demo},  the kinetic solver provides HoTs\footnote{It must be emphasized that, as shown in figure~\ref{Demo},  \eqref{eq:full_constitutive} and~\eqref{eq:getHoTs} are solved at different iteration steps, thus the NS constitutive relations cannot cancel out each other until the final steady state is reached.} to the synthetic equation~\eqref{eq:macroscopic_equation_2}, while the solution of synthetic equation, when solved to the steady-state, steers the evolution of VDFs towards the steady state. 
As has been demonstrated in various numerical simulations, this kind of treatment facilities the fast convergence in the whole range of gas rarefaction~\citep{Su2020,Liu2024,Zhang2024}. Furthermore, according to the Chapman-Enskog expansion, in the continuum limit, the HoTs are proportional to $\text{Kn}^2$~\citep{Su2020b}, such that the constitutive relation~\eqref{eq:full_constitutive} asymptotically preserves the NS limit when Kn is small.

To effectively simulate turbulence, even within the scope of the NS equations, it is essential to employ turbulence models. Given the significantly higher computational cost associated with kinetic solvers, the incorporation of turbulence models is particularly crucial.
Thanks to the explicit inclusion of the NS constitutive relations~\eqref{eq:full_constitutive}, RANS models can be seamlessly integrated into the GSIS framework. Once the turbulence viscosity from the SST model is obtained, it is superimposed into the physical viscosity present in the NS equations. Consequently, the stress and heat fluxes are formulated as follows:
\begin{equation}\label{eq:NSsigma}
\begin{aligned}
& \boldsymbol{\sigma} = 
-(\mu_{lam} + \mu_{turb})\left(\nabla{\boldsymbol{u}}+\nabla{\boldsymbol{u^\text{T}}}-\frac{2}{3}\nabla\cdot\boldsymbol{u}\textbf{I}\right) + \text{HoT}_{\bm{\sigma}},\\
& \boldsymbol{q}_{t} =  
-(\kappa_{t,lam}+\kappa_{t,turb})\nabla{T_t}
 + \text{HoT}_{\bm{q}_{t}},\quad
 \boldsymbol{q}_{r} = 
-(\kappa_{r,lam}+\kappa_{r,turb})\nabla{T_r}
 + \text{HoT}_{\bm{q}_{r}}.
\end{aligned}
\end{equation}

\begin{figure}[t]
    \centering
    {\includegraphics[trim={0 20 0 90},clip,width=0.48\textwidth]{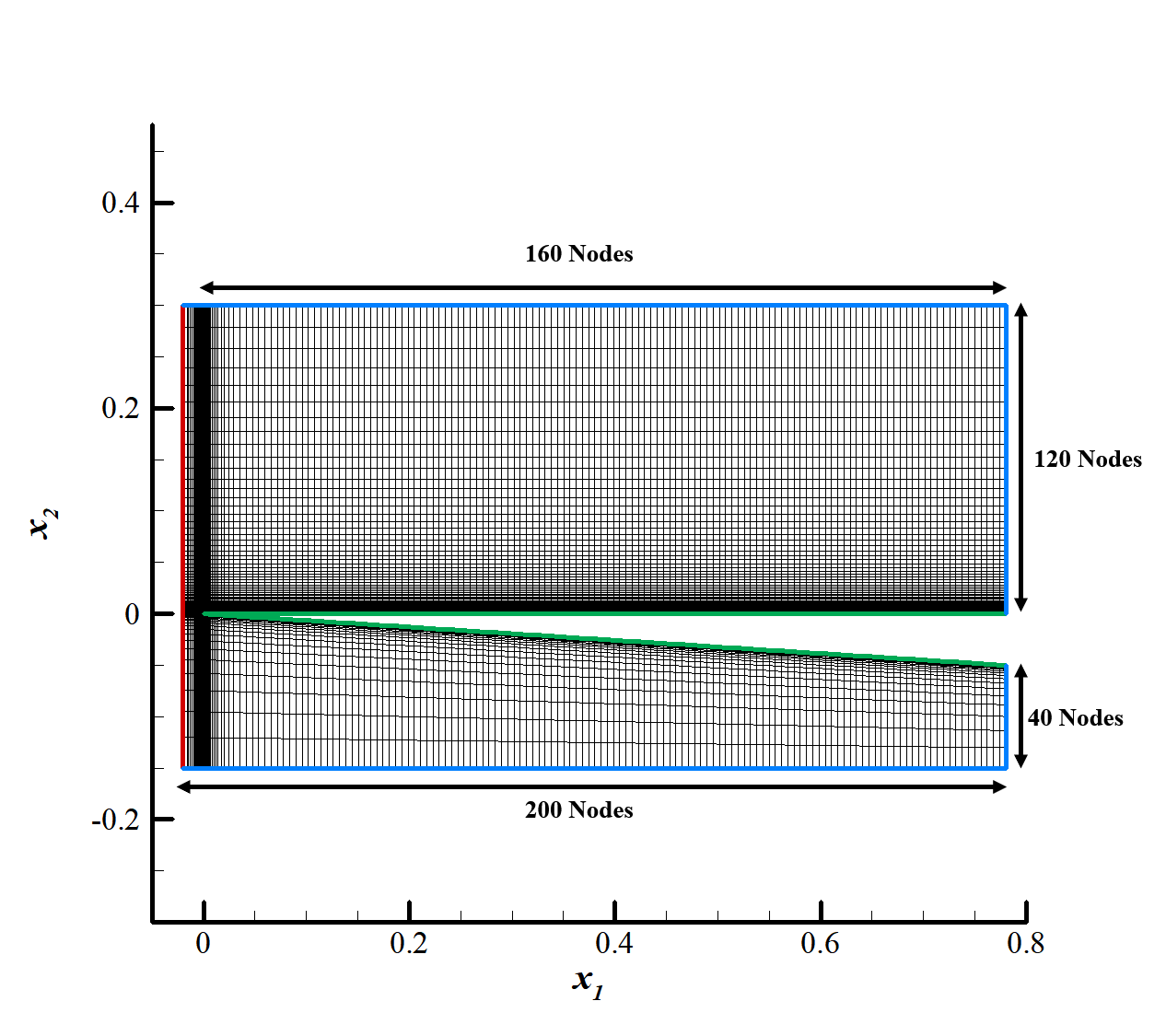}}
    {\includegraphics[trim={0 20 0 30},clip,width=0.48\textwidth]{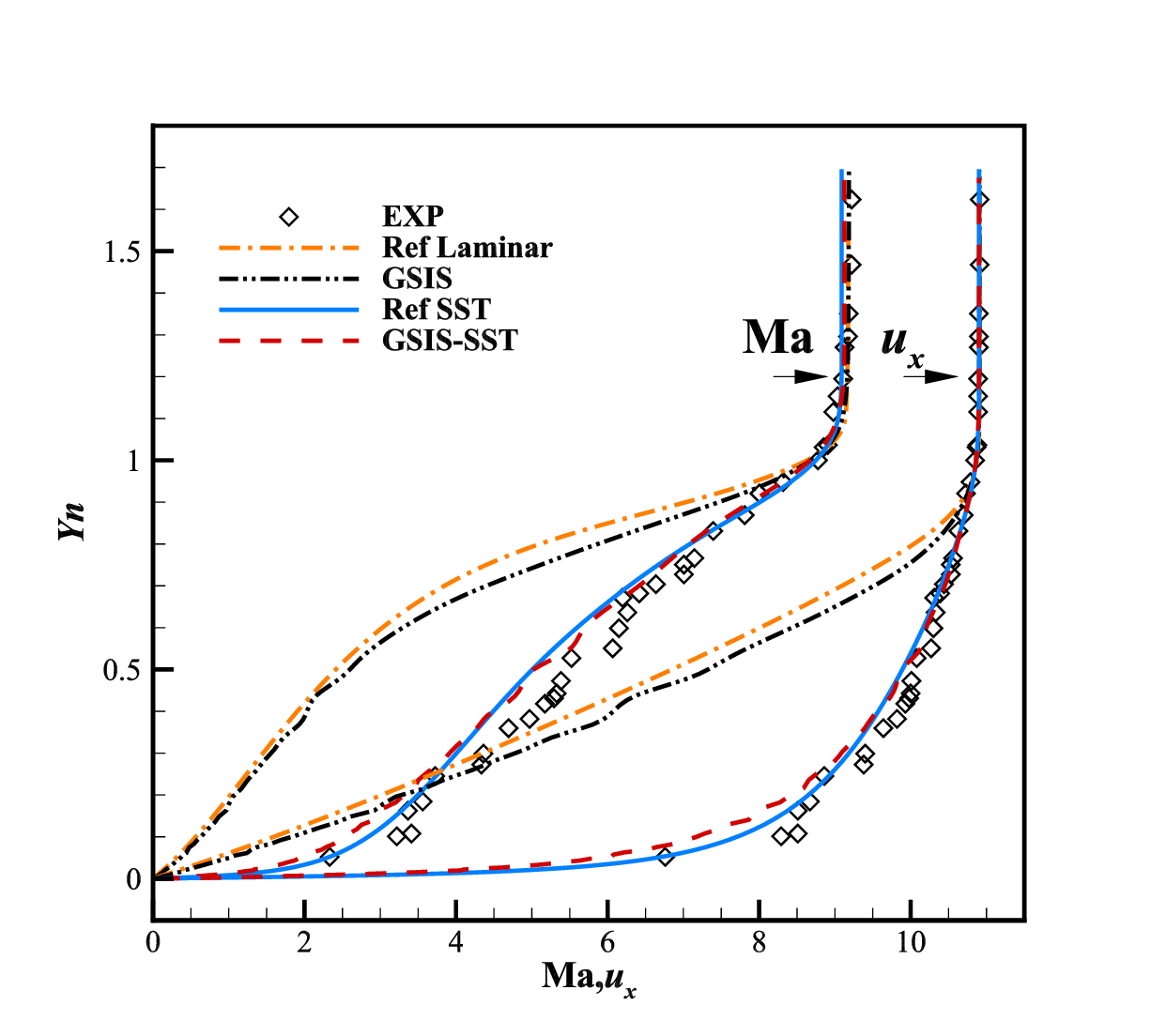}}
    \caption{
    (Left) Geometry and the computational grid (Unit: m).
    Red line on the left side stands for in-flow condition, blue lines on the top, bottom and right side are out-flow condition, while green lines are stationary isothermal wall. (Right) Velocity $u_x$ and Mach number. Experimental data are sampled at $x_1=0.076$~m on the upper flat surface~\citep{Marvin2013}. Reference solutions are obtained from conventional CFD solver. $Y_n$ is the vertical distance normalized by the boundary layer thickness measured at 99.5\% of free stream velocity. 
    }
    \label{fig:GeomGridCompressionCorner}
\end{figure}

According to the Chapman-Enskog expansion~\citep{Su2020b}, the HoTs are proportional to $\text{Kn}^2$ in the continuum flow limit, such that the GSIS-SST with the constitutive relation~\eqref{eq:NSsigma} asymptotically preserves the pure SST model when Kn is small. 
This is validated in the boundary layer problem in turbulent flows, which is a standard problem included in the NASA Turbulence modeling resource for model developing and benchmarking. As shown in figure~\ref{fig:GeomGridCompressionCorner}, we consider the hypersonic flow with Ma=9.22 passing through a wedge. 
Taking 1 m as reference length and the free flow temperature of 64.5 K as reference temperature, the Reynolds number is $4.7\times10^{7}$ and the Knudsen number is $2.91\times10^{-7}$. The solid wall temperature is 295 K. A total of 31,284 nonuniform spatial grids are used, where the height of the first layer of cells is one micron.
The two-dimensional velocity space is discretized uniformly into $300\times200$ quadrilaterals in the range of $\arrowvert \xi_x\arrowvert \times \arrowvert \xi_y \arrowvert = 30\times20$. 
Figure~\ref{fig:GeomGridCompressionCorner} shows that the boundary layer velocity and Mach number obtained from GSIS-SST fit well with the experimental data. Furthermore, the pure SST model is solved by the conventional CFD solver, which also agrees with the GSIS-SST result. Thus, the GSIS-SST's ability in turbulence modeling is confirmed. 

\begin{figure}[p]
    \centering
    \subfloat[Geometry and grid (Unit: m)]
    {\includegraphics[trim={0 20 0 40},clip,width=0.48\textwidth]{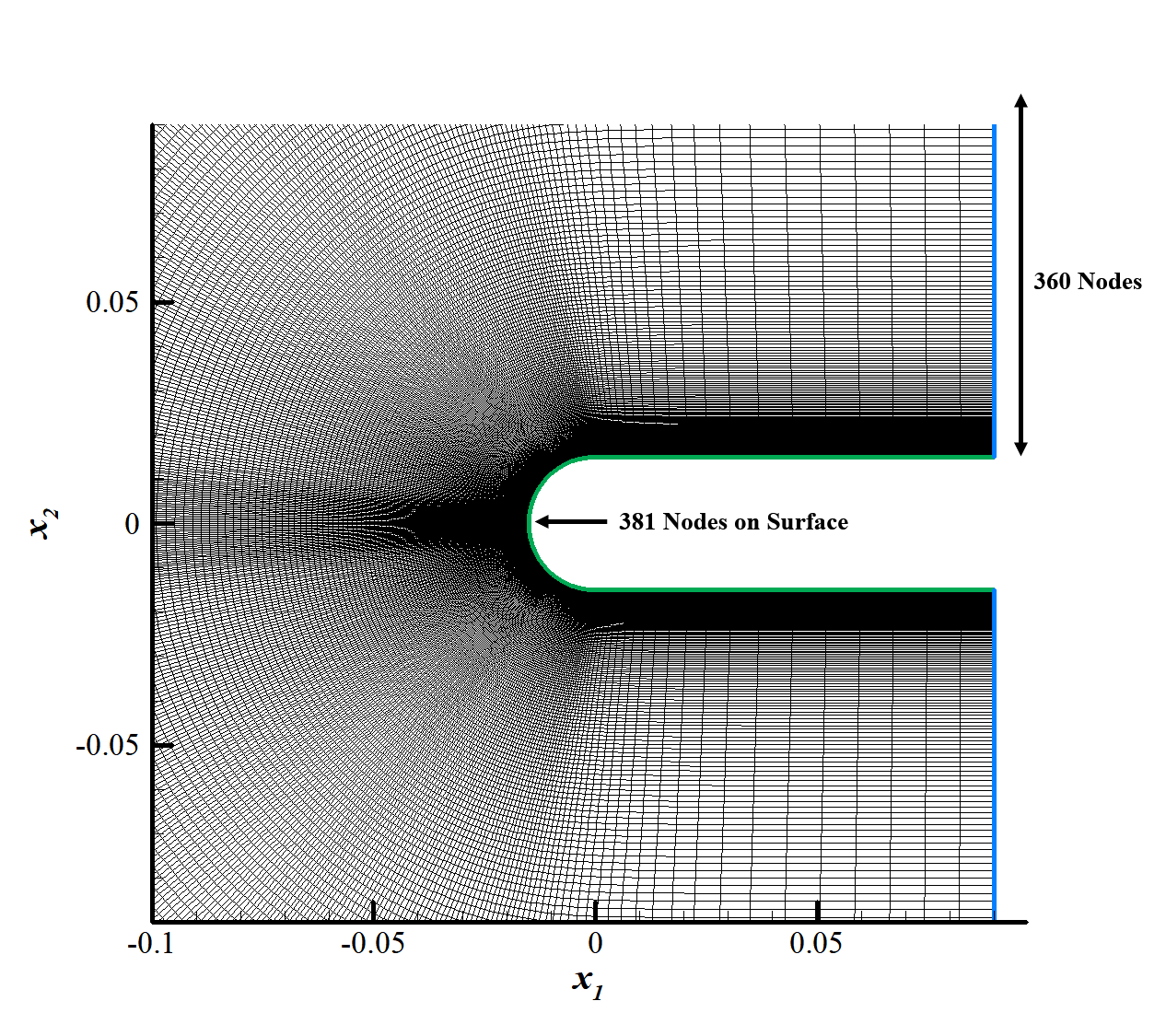}}
    \subfloat[Total temperature]
    {\includegraphics[trim={0 20 0 40},clip,width=0.48\textwidth]{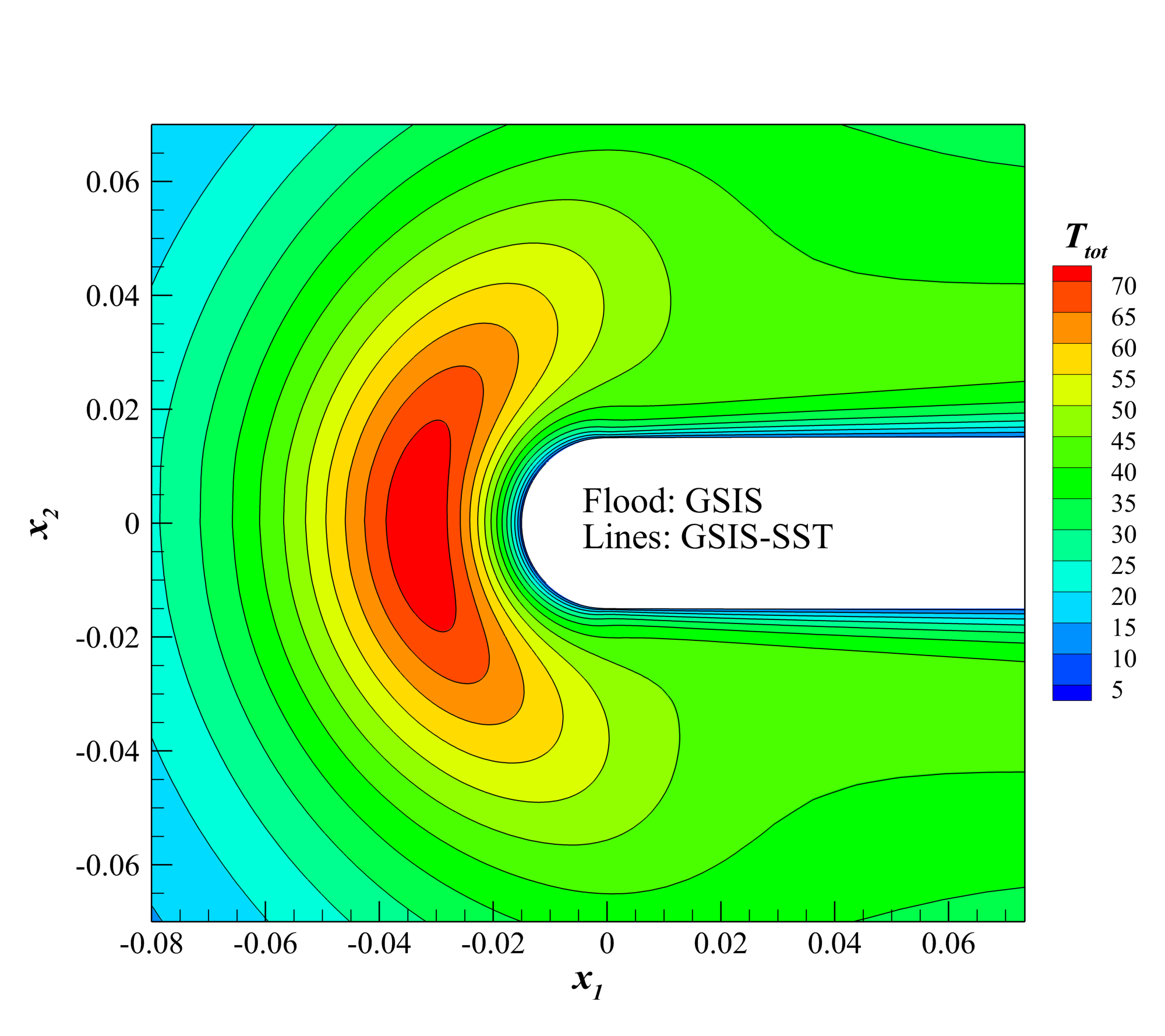}}\\
    \subfloat[Velocity and viscosity at $x_1=0.014$ m]
    {\includegraphics[trim={0 20 0 20},clip,width=0.48\textwidth]{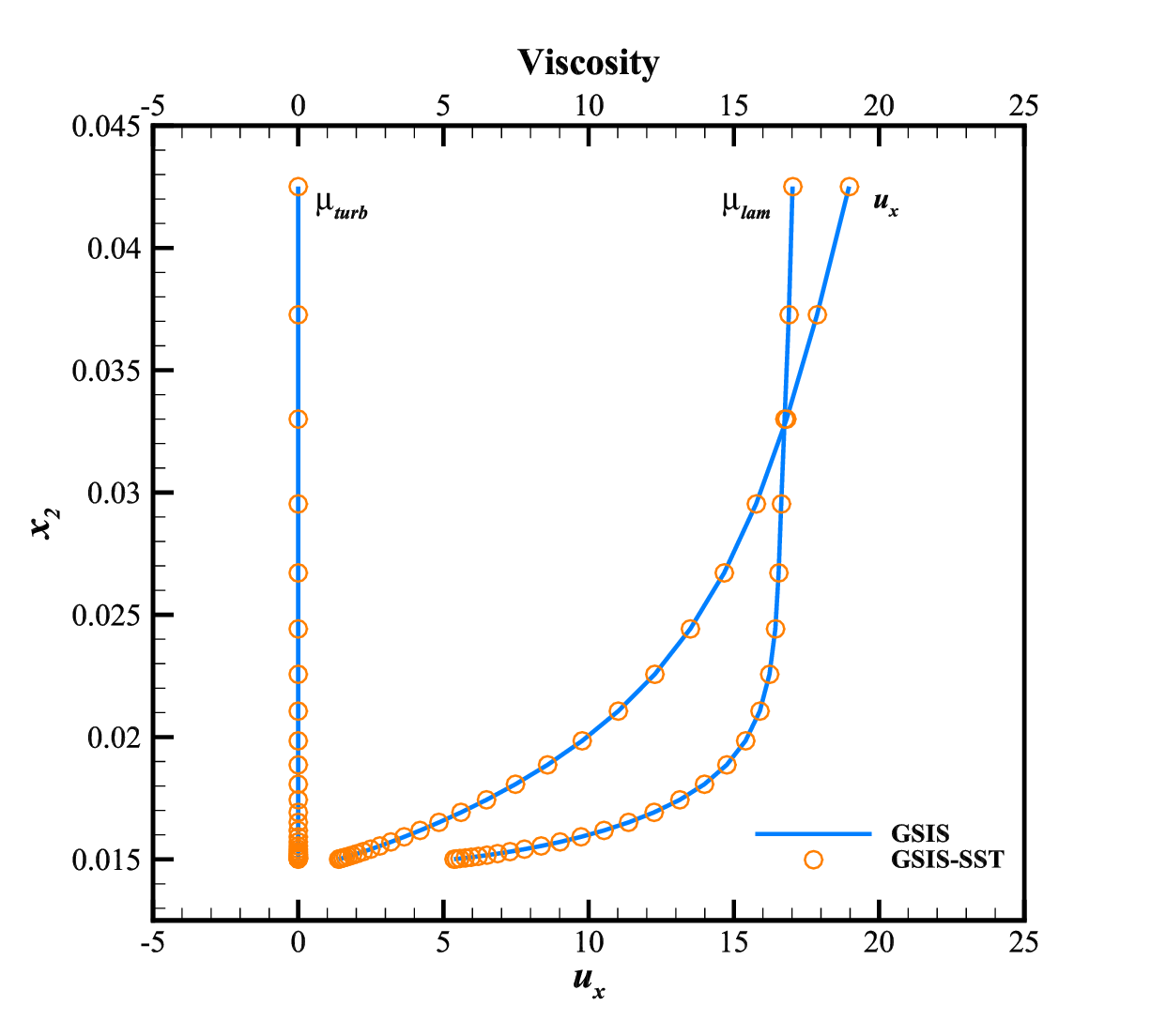}}
    \subfloat[Surface heat flux]
    {\includegraphics[trim={0 20 0 20},clip,width=0.48\textwidth]{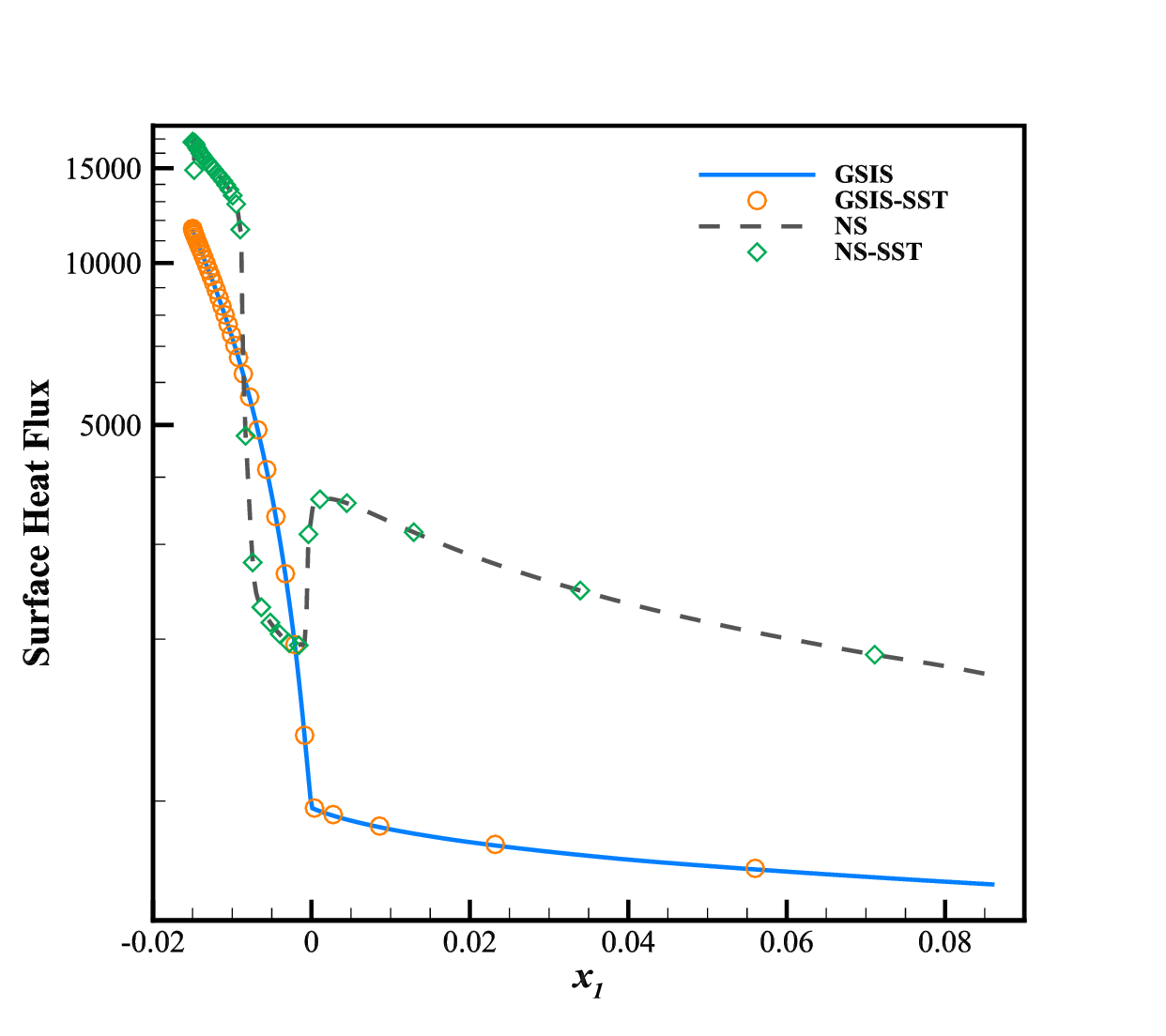}}\\
    \subfloat[Relative error]{\includegraphics[trim={0 20 0 40},clip,width=0.5\textwidth]{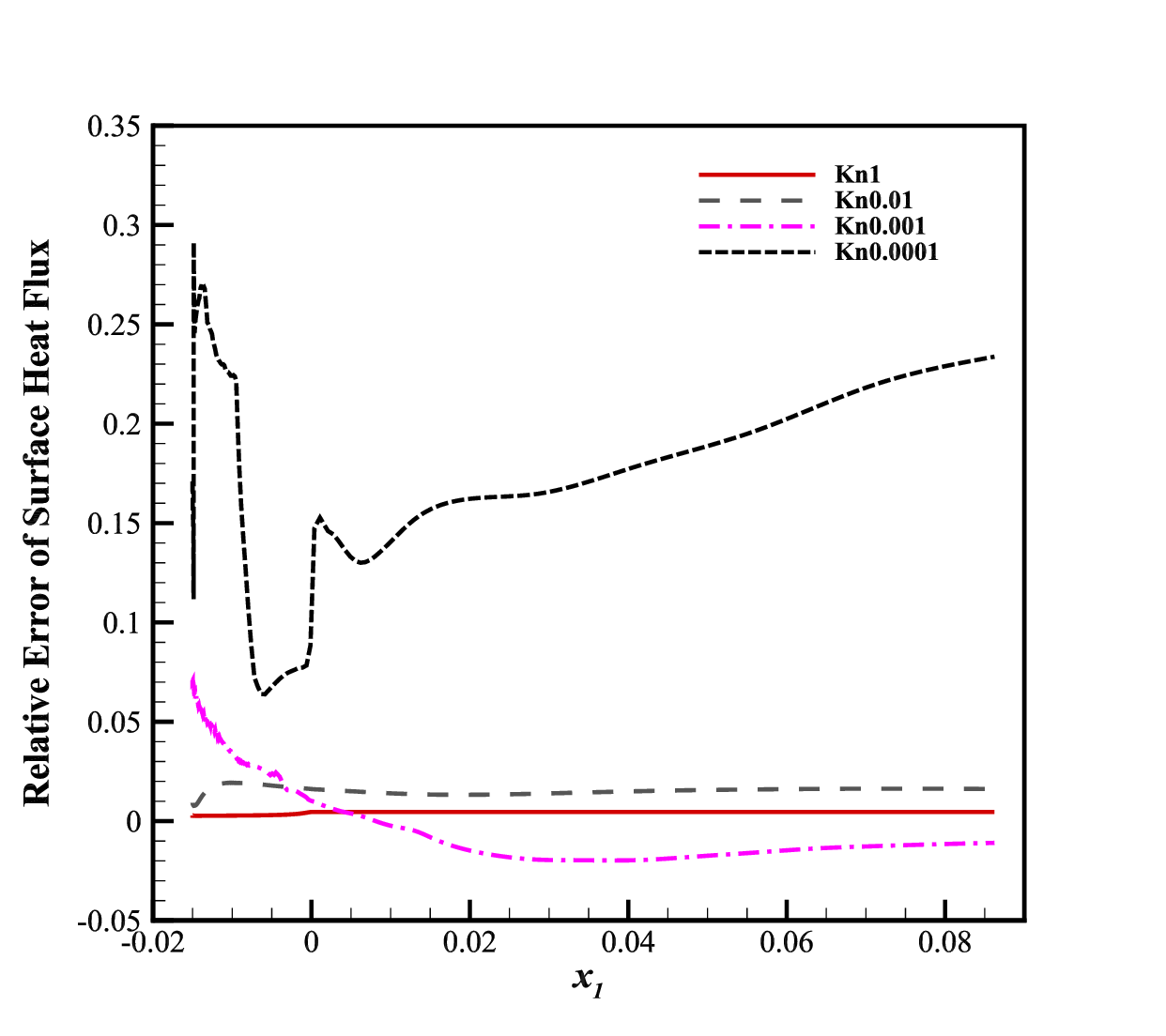}}
    \caption{Geometry and simulation results from the GSIS and GSIS-SST in the hypersonic flow of Ma=25 passing over a blunt leading edge. (a-d) Kn=1. (e) Relative difference on the total surface heat fluxes $q$, i.e., $(q_\text{GSIS-SST}/q_\text{GSIS}-1)$, when Kn=0.01, 0.001, and 0.0001. 
    } 
    \label{fig:bluntleadingedge}
\end{figure}

On the contrary, at large Kn, the turbulent viscosity $\mu_{turb}$ will be negligible compared to the physical viscosity $\mu_{lam}$, so that the GSIS-SST recovers the Boltzmann kinetic equation. 
To prove this, we simulate the hypersonic flow passing over a two-dimensional flat plate with blunt leading edge, by both the GSIS and GSIS-SST.
The computational geometry is shown in figure \ref{fig:bluntleadingedge}(a). The thickness of the plate is 0.03 m, and it is chosen as the reference length. 
The physical domain ($-0.9$ m$\le x_1 \le 0.09$ m, $|x_2| \le 0.9$ m) is discretized into 136,240 cells, with 381 nodes along and 360 normal to the model surface. The velocity space has $227\times137$ nodes.
The nitrogen gas flow of Mach number 25 is coming from the left boundary, with a temperature equal to the plate temperature, and the Knudsen number is 1 under the free streaming condition. 
Figure~\ref{fig:bluntleadingedge}(b) shows that the temperature contours, where a shock width of a few mean free path ($\lambda=0.03$) is clearly seen. The GSIS and GSIS-SST overlap perfectly with each other, enlightening that the turbulence model lose its effect in highly rarefied flows. The velocity $u_x$ at $x_1=0.014$~m is plotted in figure \ref{fig:bluntleadingedge}(c), which exhibits a large slip at the solid surface and changes rather smoothly over quite a long distance, indicating the absence of turbulent boundary layer. This subfigure also shows that the turbulent viscosity $\mu_{turb}$ is nearly zero, which is much smaller than the physical viscosity $\mu_{lam}$. This is because, in such highly rarefied flow environment and small velocity gradients, the density of near-wall flow and the production of turbulent kinetic energy are both small, leading to insignificant level of $\mu_{turb}$ as per \eqref{eq:mu_t}. Therefore, the GSIS-SST  degenerates to the pure GSIS for highly rarefied flows. This is further supported by the results in figure \ref{fig:bluntleadingedge}(d), where the surface heat fluxes predicted by GSIS and GSIS-SST overlap. Meanwhile, the pure NS and NS-SST results overlap, again proving the absence of turbulent effects in such low Reynolds number (i.e., $\text{Re}\approx25$). By this way, the difference between the NS and GSIS results show that the rarefaction effects are important when $\text{Kn}=1$, where the NS constitutive relations loss validity.


Having analyzed the asymptotic behaviors of GSIS-SST well into the turbulent and rarefied regimes, respectively, it is interesting to test its performance at intermediate Knudsen numbers. Since the pure GSIS solves the Boltzmann kinetic equation in the steady state, it is expected that below certain Knudsen number it is unable to capture the turbulence. To this end, three additional Knudsen numbers (Kn=0.01, 0.001, and 0.0001) are simulated, and 
the relative error in surface heat fluxes are compared between the GSIS and GSIS-SST in figure~\ref{fig:bluntleadingedge}(e). It is seen that the two methods start to depart from each other sensibly at $\text{Kn}\approx0.001$, where the largest relative error reaches around 7\%, slightly higher than the engineering accepted range of 5\%. At Kn=0.0001, the relative error reaches 30\%, indicating the turbulent effects are significant. This example shows that, when $\text{Re}\gtrapprox 25,000$ the turbulence model is effectively on, while when $\text{Re} \lessapprox 25,000$ the SST model is effectively off in the GSIS-SST.

The above three numerical examples show that the GSIS-SST provides reasonable modeling of multiscale flows spanning from turbulent to rarefied conditions. 


\section{Opposing Jet in Rarefied Environment}

We now employ the proposed GSIS-SST to identify engineering scenarios where turbulence and rarefaction phenomena coexist. We believe these conditions are most likely to arise during trans-atmospheric hypersonic flight. 
To achieve thermal protection and maneuverability of the aircraft, active jet technology is used to regulate aerodynamics, e.g., experiments have demonstrated the significant heat and drag reduction of the opposing jet in supersonic flows~\citep{Hayashi2003,Daso2009}. 
Numerical researches on opposing jet are abundant, where
turbulence models are always present, especially the $k$-$\omega$ SST model~\citep{Ou2018}. 
On the other hand, kinetic solvers are used to study the rarefied jet flows, e.g., the opposing jet at altitude over 70 km is simulated by DSMC~\citep{Raeisi2019,Guo2024}, and the micro-scale jets into quiescent space and vacuum plume are solved by the hybrid NS-DSMC solver~\citep{Virgile2022,Liu2024a}, where only laminar flows are considered. Here, we apply the GSIS-SST model to investigate the interaction of turbulent and rarefied flows, for the first time.

\subsection{Numerical configurations}

The physical model is shown in figure \ref{fig:bluntleadingedge}(a), except that now a 0.004 m wide slot is placed at the head of the blunt leading edge, from which pressurized nitrogen is injected towards the incoming free stream flow of Mach number 25. 
The jet is characterized by the  pressure ratio $P_{ratio}$ between the jet flow and the post-shock value of free flow\footnote{A constant pressure ratio could ensure a quite similarly located jet flow field, even if all free flow conditions change significantly \citep{Tian2023}.}:
\begin{equation}\label{pressure_ratio_jet}
P_{ratio} = \frac{p_{j,0}}{p_{f,s,0}}=2.5,
\end{equation}
where $p_{j,0}$ is the stagnation pressure of the jet, and $p_{f,s,0}$ is the stagnation pressure of the incoming shock in the downstream (calculated with the normal-shock relation based on free stream quantities). 
The jet has a static temperature of 250 K and an exit Mach number of 1.

Since the turbulence model is used in GSIS-SST, the initial values of the turbulent kinetic energy $k$ and dissipation frequency $\omega$ should be specified for the free flow boundary. Here, a turbulence intensity $I_t=0.3\%$ and a turbulent to laminar viscosity ratio $\mu_{r}=\mu_{turb}/\mu_{lam}=15$ are assumed, while a characteristic length-scale $l_t=0.118$ (nondimensionalized) is provided to the jet, 
and $I_t=3\%$ for the jet. Then $k$ and $\omega$ are derived as $k=1.5(I_t \Vert \bm{u} \Vert)^2$ and $\omega=\sqrt{k}/l_t$. 



We consider the  hypersonic flow at altitudes are 80 km and 62 km, where the Knudsen numbers are 0.125 and 0.01, respectively, when the reference length is chosen as the diameter of the model leading edge.  
The spatial grid is the same as that in figure \ref{fig:bluntleadingedge}(a), except that 39 nodes out of the 381 nodes on the surface are allocated to jet boundary.
The locally refined 2D velocity space is defined in the range of $-60 \le \xi_x \le 80$ (227 nodes) and $-65 \le \xi_y \le 65$ (137 nodes), and is cut into 30,736 cells.
Grid independence is achieved.

\subsection{Jet flow field under rarefied environment}\label{sec:JetCases}

We use the GSIS and GSIS-SST to investigate the role of turbulent jet in rarefied environment. The NS and NS-SST equations are not used since they do not describe the rarefaction effects, e.g., see figure~\ref{fig:bluntleadingedge}(d). 

\begin{figure}[t]
    \centering
   \subfloat[Kn=0.01]{ \includegraphics[trim={0 50 200 20},clip,width=0.48\textwidth]{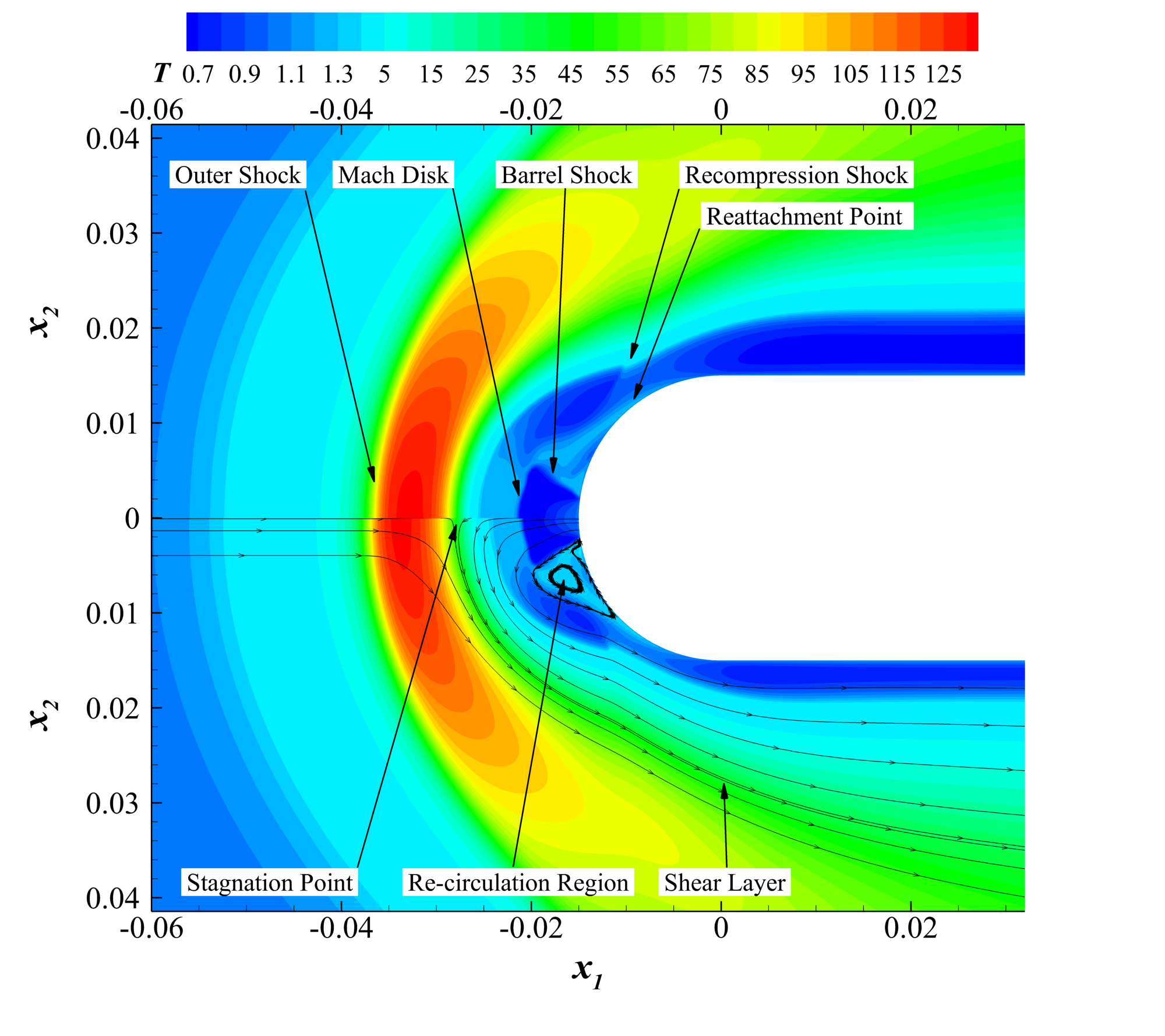}}
    \subfloat[Kn=0.125]{ \includegraphics[trim={0 50 200 20},clip,width=0.48\textwidth]{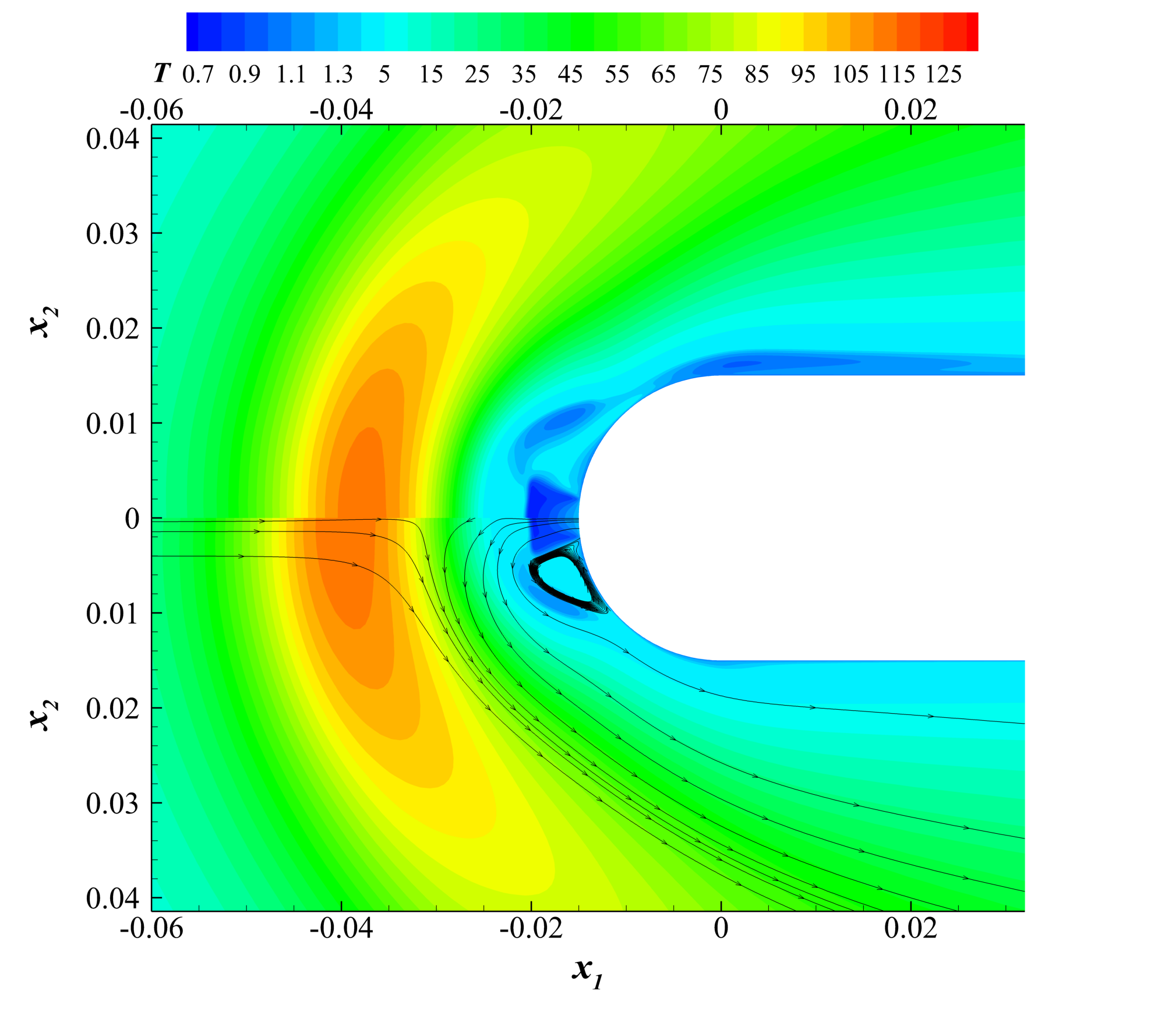}}
    \caption{Comparison of temperature contour from GSIS (upper half) and GSIS-SST (lower half). 
    }
    \label{fig:KnContourCP}
\end{figure}

Figure \ref{fig:KnContourCP} depicts the temperature contours when Kn=0.01 and 0.125.
The under-expanded jet is injected into the flow field from the nozzle at $x_1=-0.0150$ m and is confined by the barrel shock. It expands and accelerates vigorously until it encounters the free flow, resulting in a pronounced Mach disk. The Mach disk acts to compress and decelerate the now over-expanded jet flow. Beyond the Mach disk, the jet flow achieves a local equilibrium with the post-shock free flow, subsequently flowing laterally and in reverse, reattaching to the model surface. A recirculation zone is present between the barrel shock, the reversed jet flow, and the model surface, characterized by closed streamlines in the temperature contour. This zone arises from two separation phenomena: the first caused by the jet's expansion at the nozzle, and the second induced by the reversed jet flow’s reattachment to the model surface. The reattachment stimulates the development of a recompression shock, and the change in surface curvature from the blunt to the flat section leads to flow expansion near $x_1=0$ m.
The jet displaces the high-temperature post-shock region, creating a low-temperature sheath that exists between the hot outer shock and the model. This cool sheath is roughly delineated by the streamlines emanating from the jet exit. Consequently, this cool sheath is also known as the jet-controlled area.

When Kn increases, the outer shock gets thicker due to the increased non-equilibrium effect. In the jet-controlled area, flow structures are sharp and clear. 
The position of Mach disk moves slightly closer to the jet exit. 
The shape of Mach disk becomes flatter when Kn gets higher. When Kn=0.01, the post-shock stagnation pressure of the free flow is actually lower than lateral positions, since the oblique outer shock is weaker laterally, thus a curved Mach disk is formed. However, when Kn=0.125, the rarefaction effect weakens the outer shock and reduces such lateral difference, thus producing a flatter Mach disk. 



\begin{figure}[t]
    \centering
   {\includegraphics[trim={0 50 0 20},clip,width=0.48\textwidth]{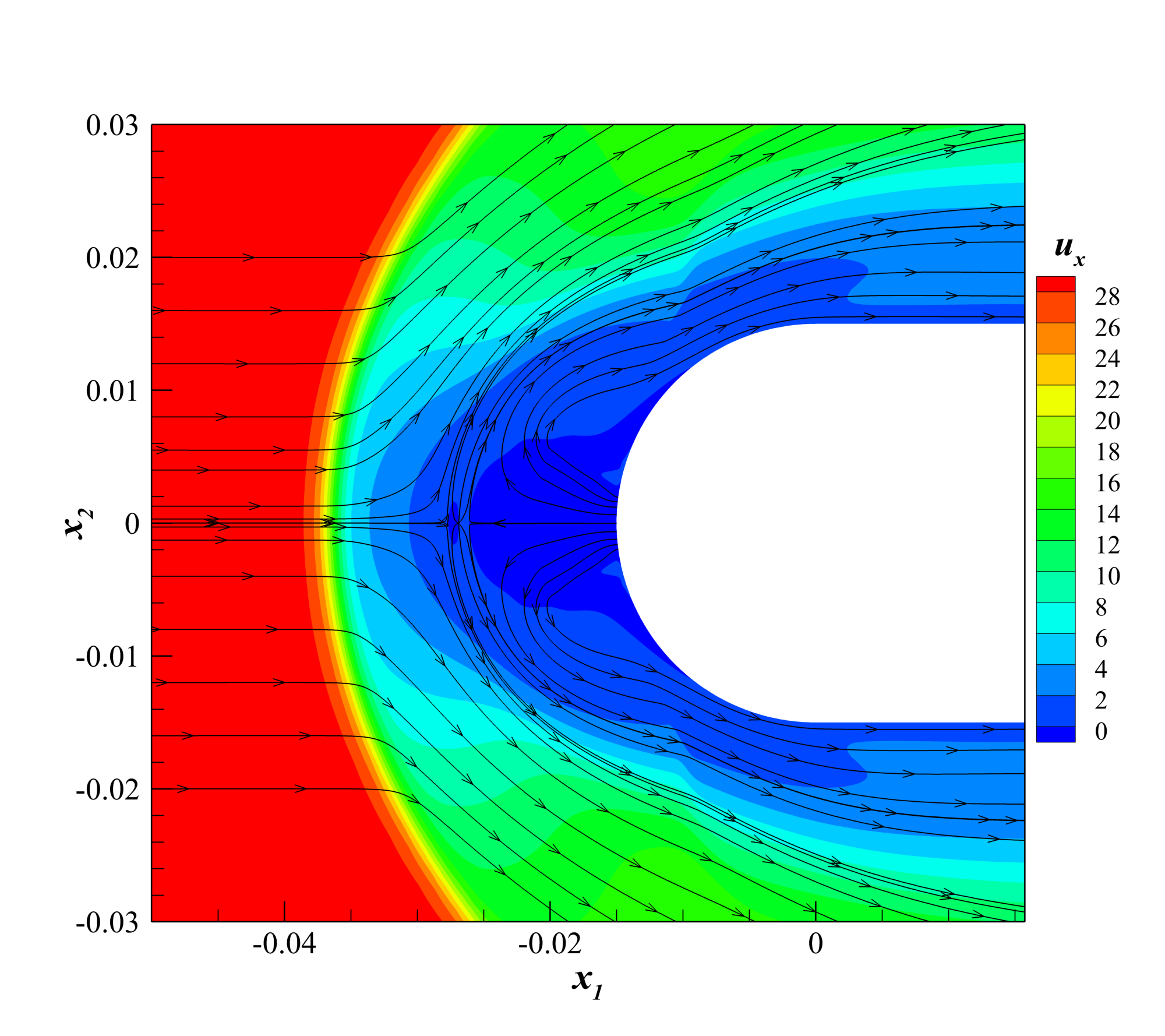}}
    {\includegraphics[trim={0 20 0 20},clip,width=0.48\textwidth]{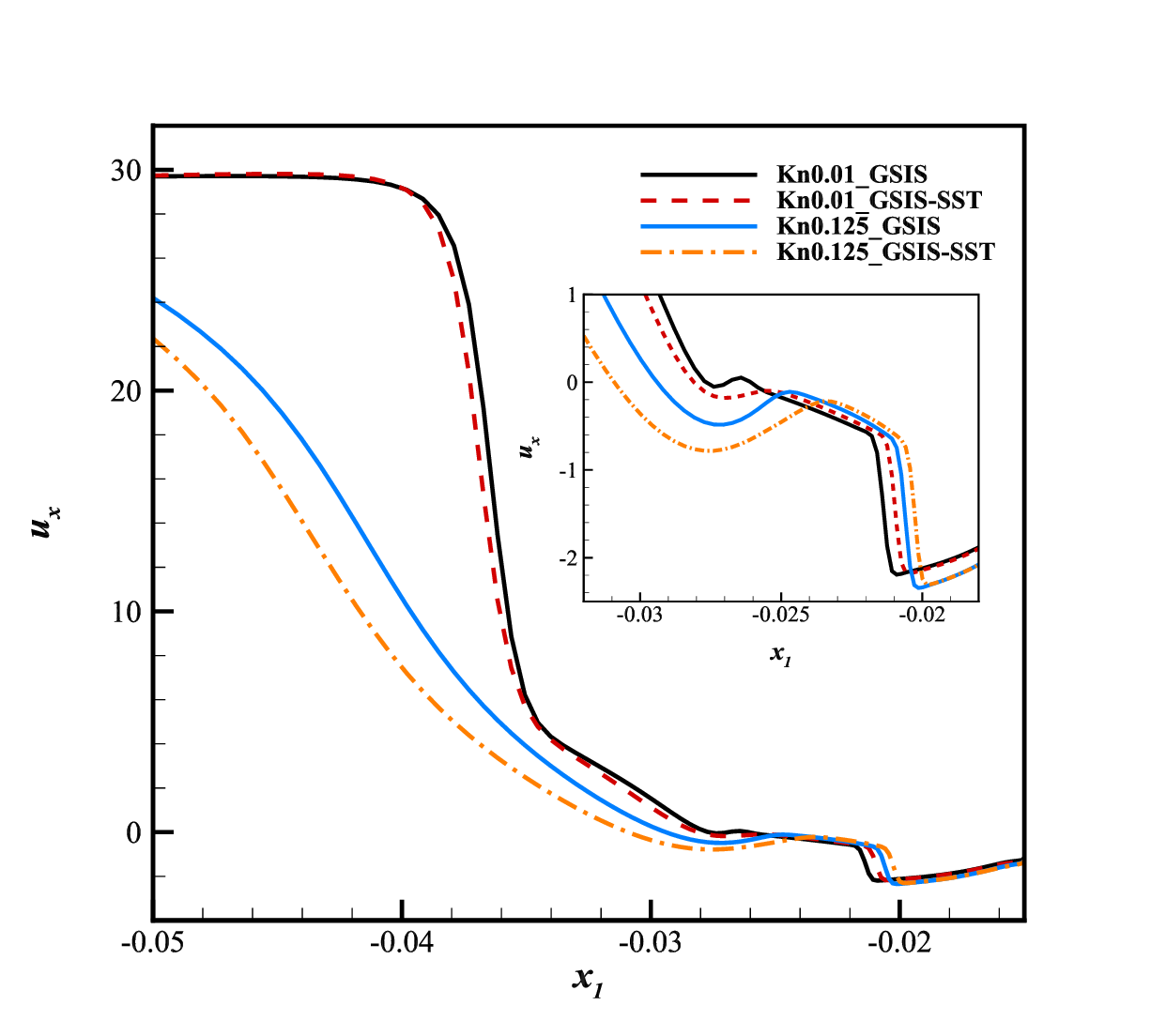}}
    \caption{(Left) Contour of horizontal velocity with streamlines from GSIS at Kn=0.01. (Right) Horizontal velocity along the centerline at Kn=0.01 and 0.125. Inset: zoomed-in figures to show the diffusion layer.
    }
    \label{fig:KnUxStreamPlots}
\end{figure}

The flow velocity in figure~\ref{fig:KnUxStreamPlots} reveals another change of flow pattern under rarefied environment: production of an additional inflection point (around $x_1=-0.027$ m) along the center line. 
The free flow coming from the left experiences a drop in $u_x$, while the under-expanded jet flow coming from $x_1=-0.015$ m experiences an increase of magnitude in $u_x$. Normally, this change of velocity should be monotonic. 
However, figure~\ref{fig:KnUxStreamPlots}(b) shows that $u_x$ follows a decrease-increase-decrease process. Such effect becomes stronger when Kn increases. We believe that it is a consequence of the rarefied environment: the low number density free flow and jet flow lead to more free space among molecules and less frequent collisions, shaping an apparent and thick diffusion layer. Thus, this counter-intuitive change is the outcome of the microscopic mixing process, in which molecules from both direction flow into each other for a certain range without sufficient collision. In contrast, in continuum regime such layer is too thin to be noticed. 

On the influence of turbulence, GSIS and GSIS-SST contours and plots in figures \ref{fig:KnContourCP} and \ref{fig:KnUxStreamPlots} look alike in general shape under the same Kn condition. 
Since the introduced turbulence functions as an extra viscosity, structures with steep changes are generally smoothed in GSIS-SST. Another alternation is the shift of Mach disk, which stands slightly closer to the model in GSIS-SST. And due to the enhanced mixing of the jet and free flow, the thickness of low temperature layer normal to the wall shrinks, see blue regions around the model figure \ref{fig:KnContourCP}. These changes finally build up the difference on surface quantities distributions, as will be quantitatively analyzed below.




\subsection{Interaction between turbulent and rarefied flows}


The boundary layer is examined to study the interaction between the jet, the rarefaction effect and the turbulence in the near wall region. Furthermore, the turbulent viscosity $\mu_{turb}$ is calculated to understand how does the turbulence shape the flow field.

\begin{figure}[t]
    \centering
    \subfloat[Kn=0.01]{\includegraphics[trim={0 10 30 60},clip,width=0.48\textwidth]{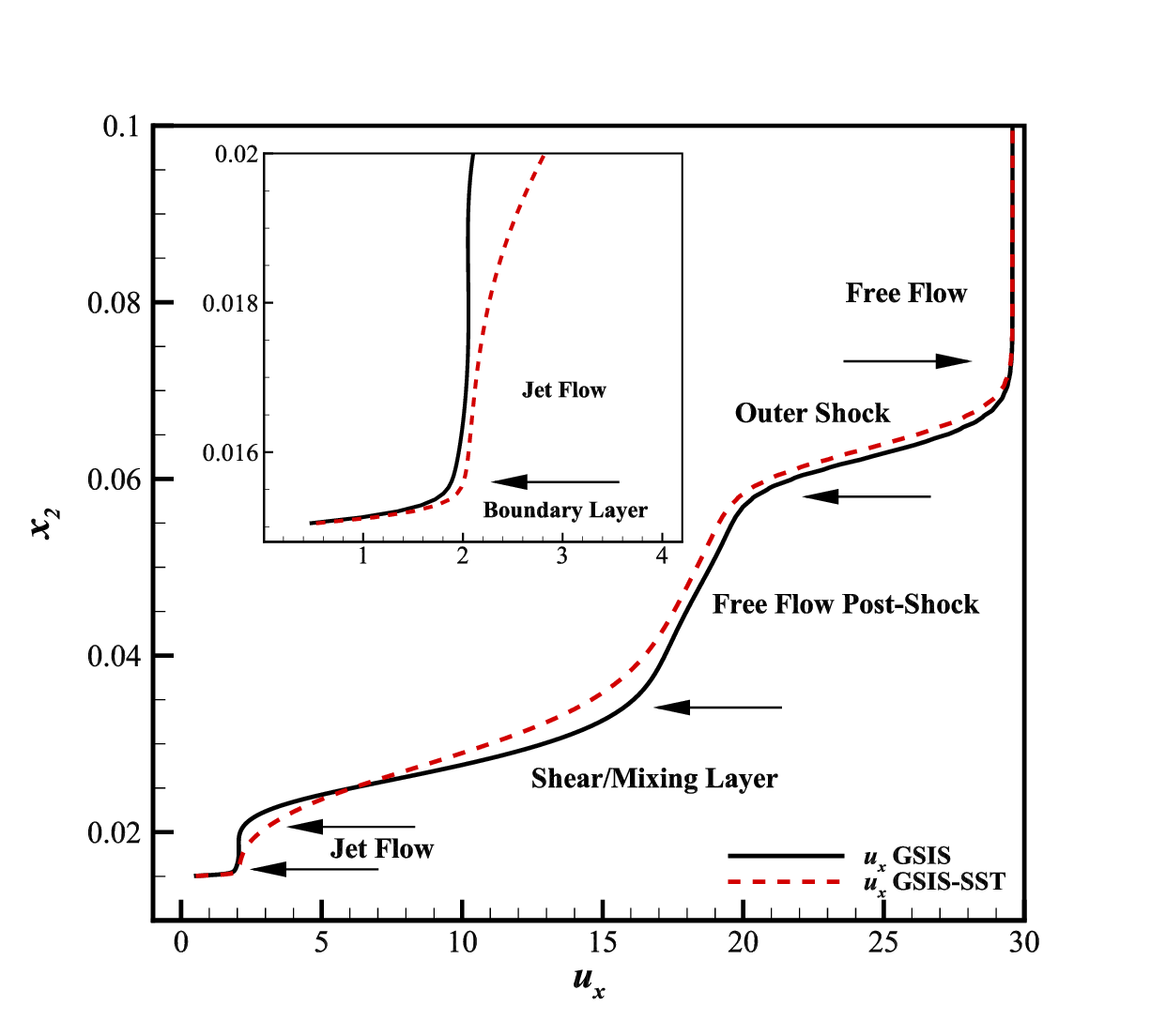}}
    \subfloat[Kn=0.01]{\includegraphics[trim={0 10 30 60},clip,width=0.48\textwidth]{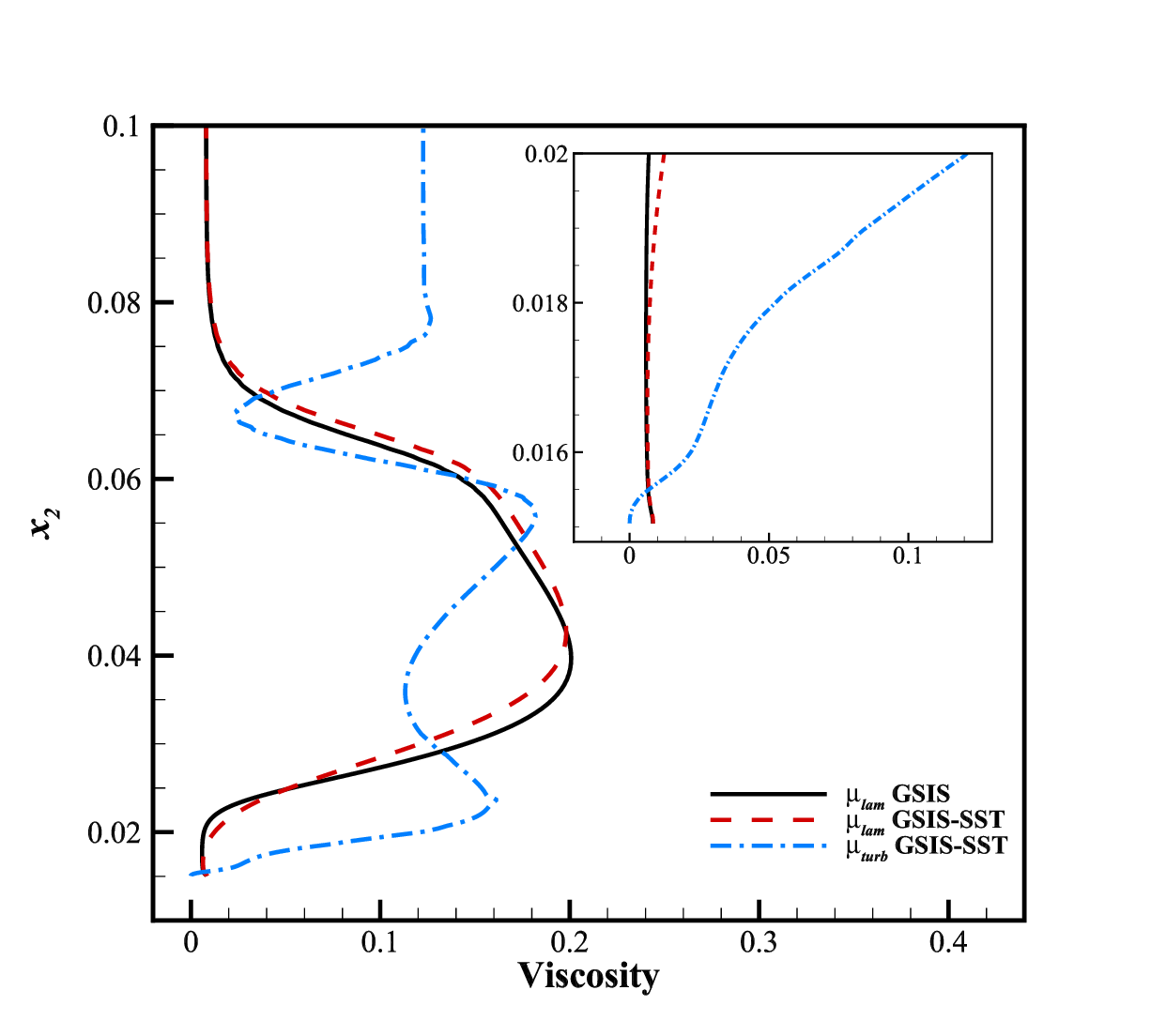}}\\
    \subfloat[Kn=0.125]{\includegraphics[trim={0 10 30 60},clip,width=0.48\textwidth]{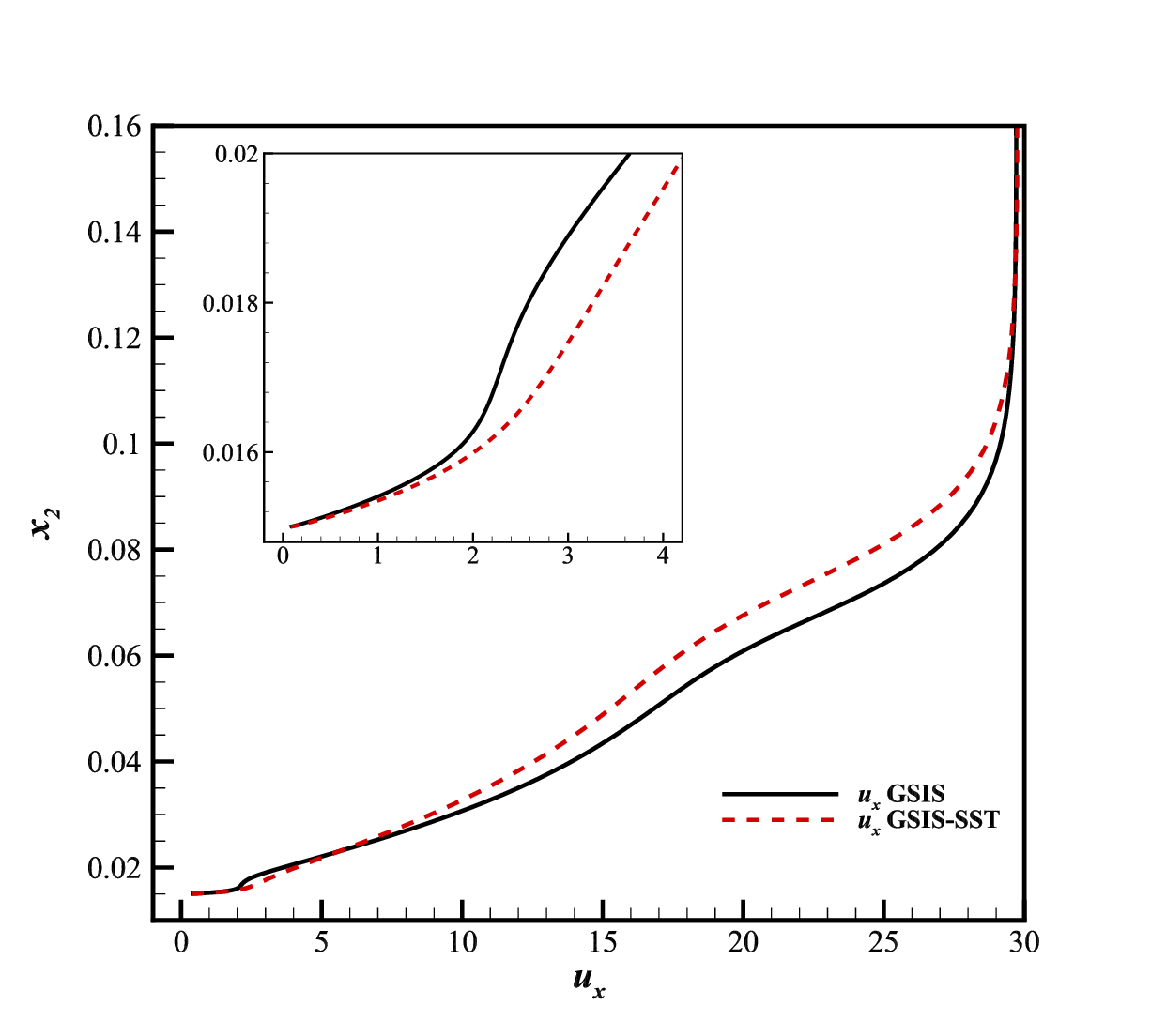}}
    \subfloat[Kn=0.125]{\includegraphics[trim={0 10 30 60},clip,width=0.48\textwidth]{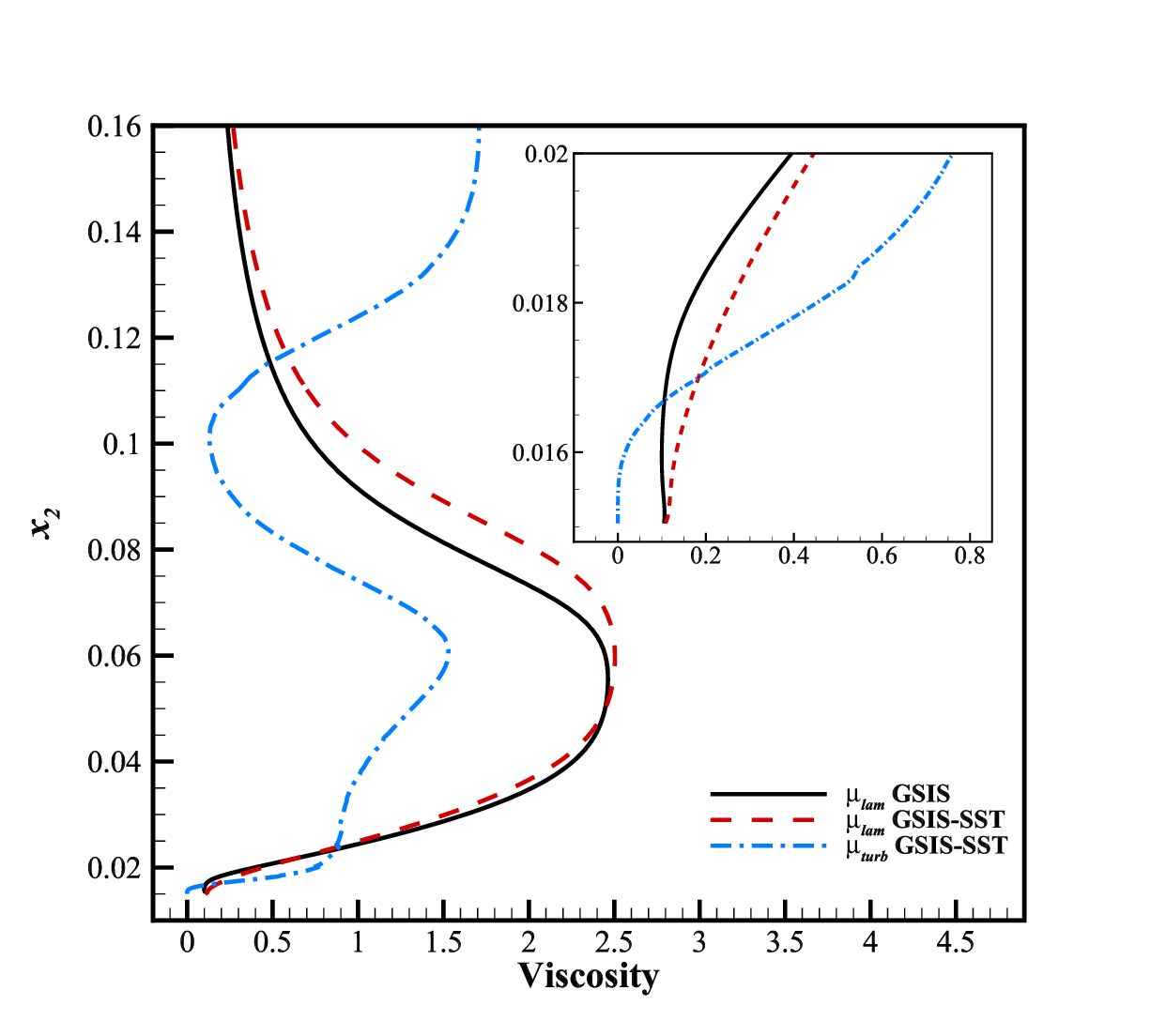}}
    \caption{Profile of horizontal velocity and viscosity at $x_1=0.006$~m.
    Inset: velocity and viscosity in the vicinity of solid wall.
    }
    \label{fig:KnSweep_VelocityProfile}
\end{figure}

When Kn=0.01, figure~\ref{fig:KnSweep_VelocityProfile}(a) shows that the velocity profile from GSIS is well layered. From top to bottom, the free flow region, outer shock, free flow dominated post-shock layer, shear/mixing layer between the free and jet flow, jet flow dominated layer, and near wall boundary layer could be identified. The typical boundary layer still exist, where the flow velocity increases from 0 across the boundary layer to 2 and keeps at the jet flow velocity until $x_2\approx0.019$ m. Then, the jet flow starts to mix with free flow and $u_x$ keeps increasing. After a turning point at $x_2\approx0.035$ m, the free flow dominated layer is reached and the increasing rate of $u_x$ slows down. Through the outer shock with large gradients, $u_x$ finally reaches the undisturbed free flow velocity of 29.6 (Mach number 25). The GSIS-SST result shows generally the same pattern as the GSIS one, but the transition between layers are smoothed. Also, $u_x$ from GSIS-SST increases faster inside the boundary layer than the GSIS. 

It is the turbulence viscosity that shapes this difference. The viscosity profile from GSIS-SST at Kn=0.01 is depicted in figure~\ref{fig:KnSweep_VelocityProfile}(b), which shows that $\mu_{turb}$ increases from 0 at the wall, rises up and cross the physical viscosity $\mu_{lam}$ curve around $x_2=0.0155$ m; within this range $\mu_{turb}$ is produced due to large velocity gradient of $u_x$ in $x_2$ direction. 
The turbulence viscosity dominates over the physical viscosity in the range of $0.0155<x_2<0.0303$~m (turbulent jet flow and shear/mixing layer) as the jet flow is a source of turbulence. And it also dominates in the range of {$0.0535~\text{m}<x_2<0.0595~\text{m}$} (outer shock) due to large velocity gradients across shock. 
The high turbulence viscosity in these regions boosts the momentum and energy exchange. High speed and high temperature post-shock free flow introduce more kinetic and internal energies through intense turbulent mixing into the lower layers (shear/mixing layer and subsequently the jet flow). Therefore, abrupt transitions around $x_2\approx0.0225~\text{m}$ and $x_2\approx0.035~\text{m}$ in GSIS horizontal velocity profile is smoothed in GSIS-SST, see figure~\ref{fig:KnSweep_VelocityProfile}(a). 

When Kn goes up to 0.125, velocity profiles from GSIS and GSIS-SST become smoother (reaching free flow velocity at around $x_2=0.12$ m) than that at Kn=0.01, so that the transition of different velocity layers in figure~\ref{fig:KnSweep_VelocityProfile}(a) is now blurred in figure~\ref{fig:KnSweep_VelocityProfile}(c). 
This is the result of rarefaction effects: since the free flow is rather rarefied and the density of the jet flow is also decreased (as the pressure ratio $P_{ratio}$ in \eqref{pressure_ratio_jet} is fixed), the diffusion between different layers is intensified, disturbance travels further into each other, leading to higher temperature and momentum around the model. In GSIS-SST, adding together the rarefaction effect and turbulence, the velocity profile is no doubt even more smoother. 
The viscosity profiles at Kn=0.125 show similar pattern as the ones at Kn=0.01, i.e., drops before and rise after the outer shock, drops again in the free flow post-shock region, rises or keeps a in the shear layer, decreases finally to 0 at wall. Yet the turbulence viscosity dominated area in GSIS-SST at Kn=0.125 is now only restricted to jet flow dominated area, i.e., when $x_2<0.0236$ m. The local peak of $\mu_{turb}$ at $x_2\approx0.055$ at Kn=0.01 still exist at $x_2\approx0.06$ in Kn=0.125, but is now smaller than the corresponding molecular viscosity. This is the result of higher level of rarefaction at Kn=0.125, since $\mu_{turb}$ is proportional to density.

\begin{figure}[t]
    \centering
    \subfloat[Kn=0.01]{\includegraphics[trim={0 50 0 20},clip,width=0.48\textwidth]{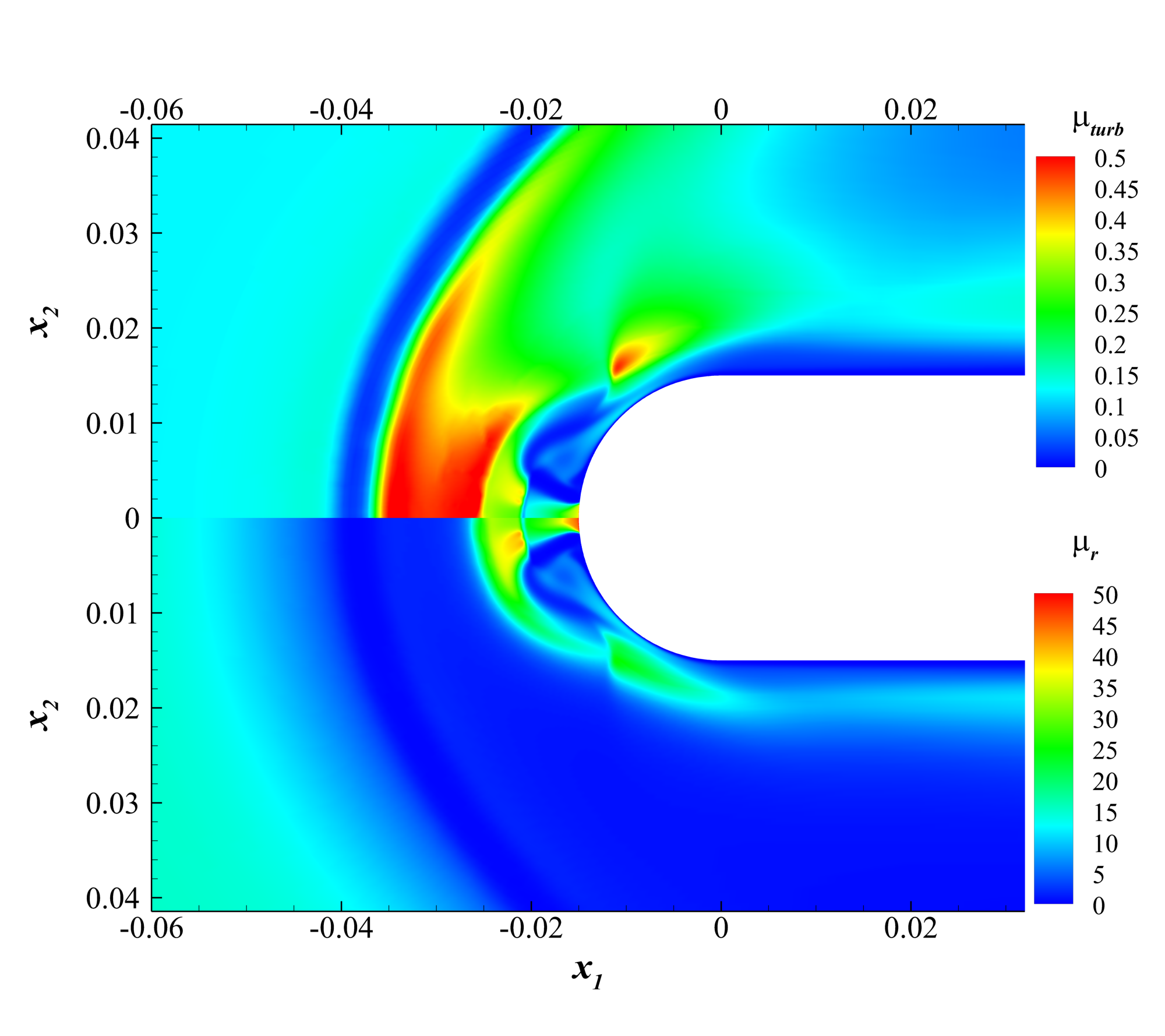}}
    \subfloat[Kn=0.125]{\includegraphics[trim={0 50 0 20},clip,width=0.48\textwidth]{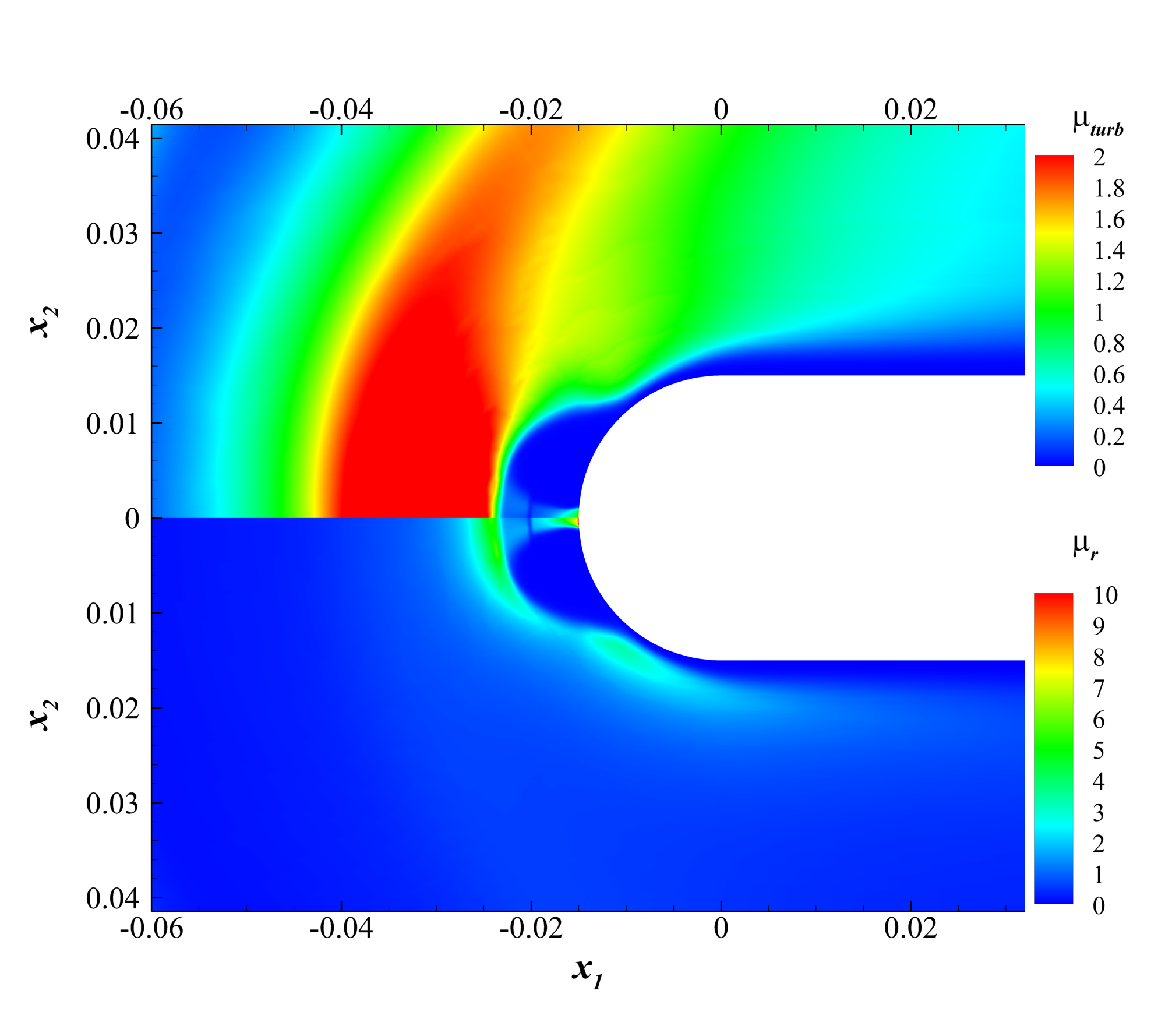}}
    \caption{
    Contour of turbulent viscosity $\mu_{turb}$ (top half) and the turbulent to physical viscosity ratio $\mu_r$ from GSIS-SST (bottom half). The initial jet, the post Mach disk region and the reattachment region are the main source of turbulence effect.  
    }
    \label{fig:KnSweepMuRatioCP}
\end{figure}

Figure \ref{fig:KnSweepMuRatioCP} shows the ratio between the turbulent and physical viscosities in GSIS-SST.
When Kn=0.01, distributions of $\mu_{turb}$ exhibit a pronounced concentration towards the Mach disk, the outer shock (around $x_1=-0.035~\text{m}$) and the recompression shock (around $x_1=-0.01~\text{m}$). These structures are associated with high velocity gradients which promotes production of turbulent kinetic energy and hence $\mu_{turb}$. Unlike $\mu_{turb}$, the viscosity ratio $\mu_r$ has only high values in the cooler jet-controlled area (refer to figure~\ref{fig:KnContourCP}, region between model surface and shear layer): near jet exit, left to the Mach disk and the recompression shock. Since the jet temperature is much lower and hence lower $\mu_{lam}$, $\mu_{turb}$ could be dominant in these regions.
When Kn=0.125, $\mu_{turb}$ still shows high concentration in the region between the outer shock and the Mach disk, the recompression shock contributes now less to the formation of $\mu_{turb}$ compared to the Kn=0.01 case. The viscosity ratio $\mu_r$ at Kn=0.125 shows similar distribution as that at Kn=0.01, yet the high $\mu_r$ shrinks at Kn=0.125.

It is clear that the high $\mu_{turb}$ (both at Kn=0.01 and Kn=0.125) behind the outer shock does not turn into actual effect, as the $\mu_r$ there is quite low. This is because the high post-shock temperature largely rises $\mu_{lam}$, diluting the influence from $\mu_{turb}$.

\subsection{Changes in surface quantities}

\begin{figure}[t]
     \centering
\subfloat[]{ 
\includegraphics[width=0.48\textwidth]{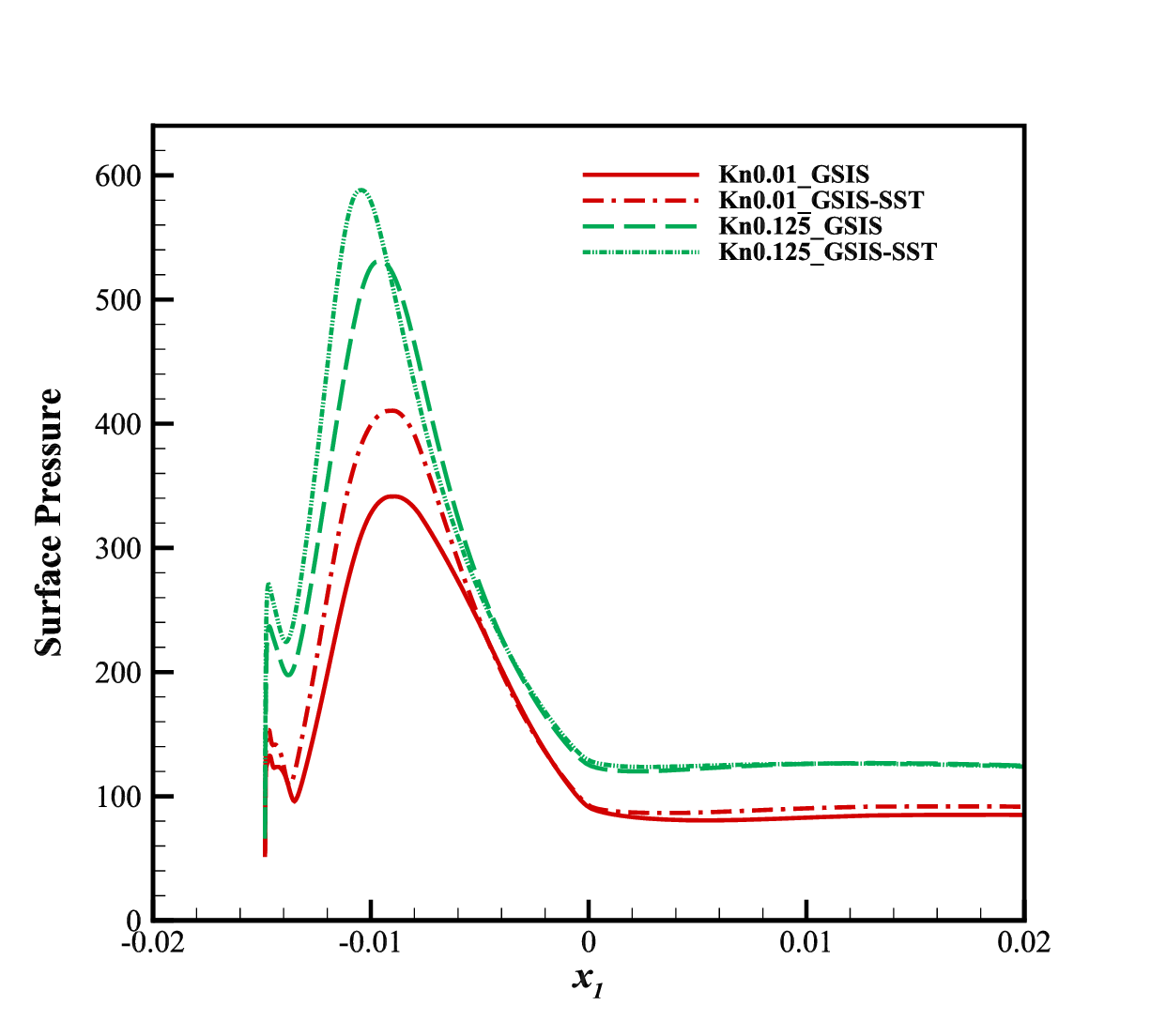}}
\subfloat[]{ \includegraphics[width=0.48\textwidth]{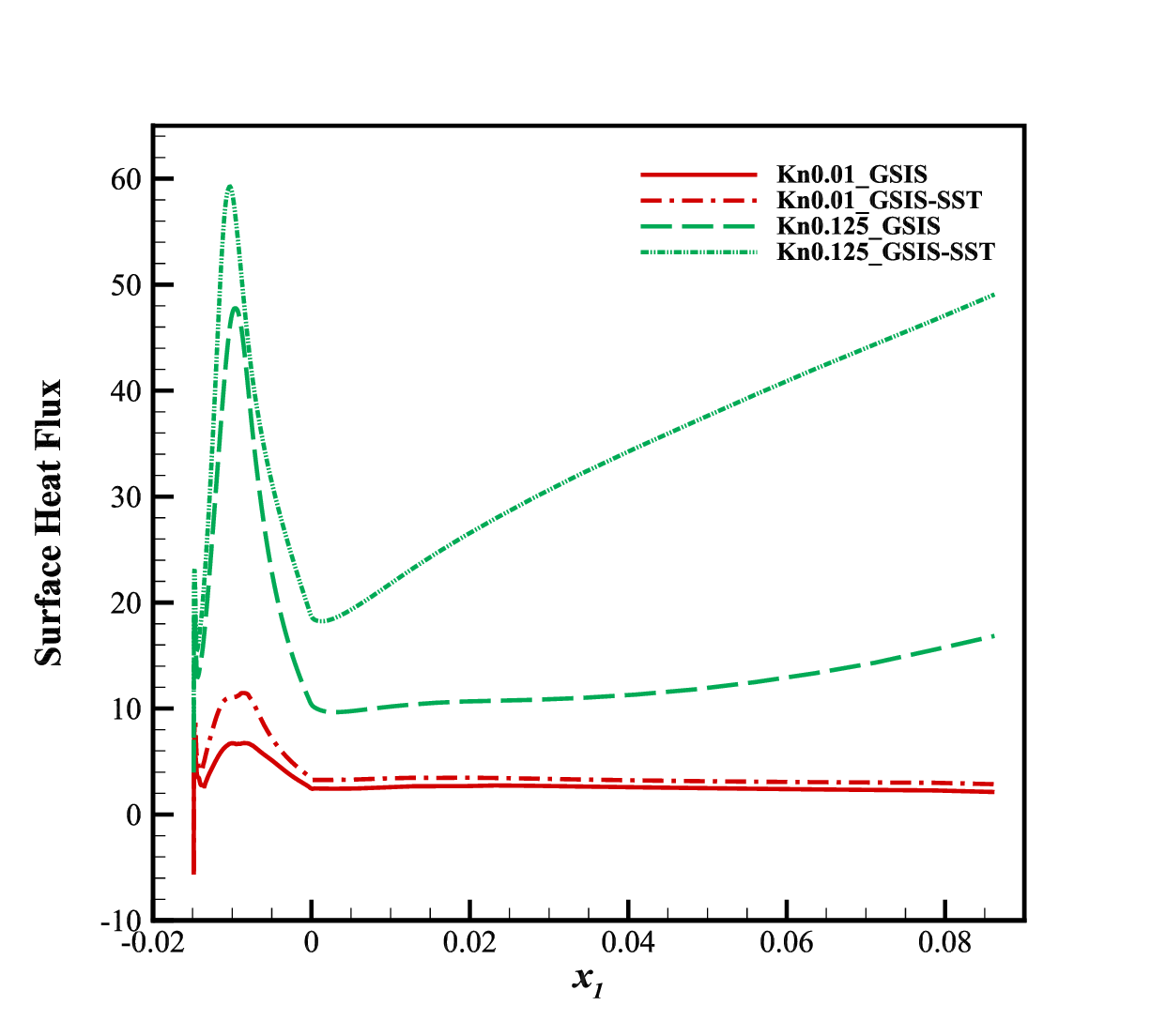}}
     \caption{(a) The surface pressure. Results from GSIS and GSIS-SST converge after $x_1=0.02$ m, and hence are not shown.
     (b) Surface heat flux.
     }
          \label{fig:JetPressQtwPlot}
 \end{figure}

Surface pressure and heat flux are of great engineering significance, and the differences between the GSIS and GSIS-SST are analyzed here. 
Figure~\ref{fig:JetPressQtwPlot}(a) shows that, when the Knudsen number is fixed, the surface pressure is small around the jet nozzle, grows through the barrel shock and drops in the re-circulation region, increases again to peak value at reattachment point, and decreases almost monotonically after the peak. 
When the Knudsen number increases from 0.01 to 0.125, the pressure profile barely changes, but its magnitude increases. Comparison between the GSIS and GSIS-SST results show that the turbulence model increases the surface pressure, e.g., surface pressure from GSIS-SST is higher than that of GSIS by 20\% and 11\% at Kn=0.01 and 0.125, respectively.



Figure~\ref{fig:JetPressQtwPlot}(b) shows  
the distributions of surface heat flux $q$. When Kn=0.01, $q$ obtained from the GSIS is small near the nozzle exit, and increases to peak value at the reattachment point (around $x_1=0.0125~\text{m}$). After the peak, $q$ drops fast along the surface of the blunt leading edge, and then stays nearly constant along the flat plate. When Kn=0.125, the variation of $q$ curve, obtained from the GSIS, is similar to that at Kn=0.01, except that when $x_1>0~\text{m}$ the surface heat flux starts to climb again, first gradually and then accelerates (from $x_1\approx0.05~\text{m}$) when approaching the end of the model. 

\begin{figure}[t]
     \centering
     \includegraphics[width=0.48\textwidth]{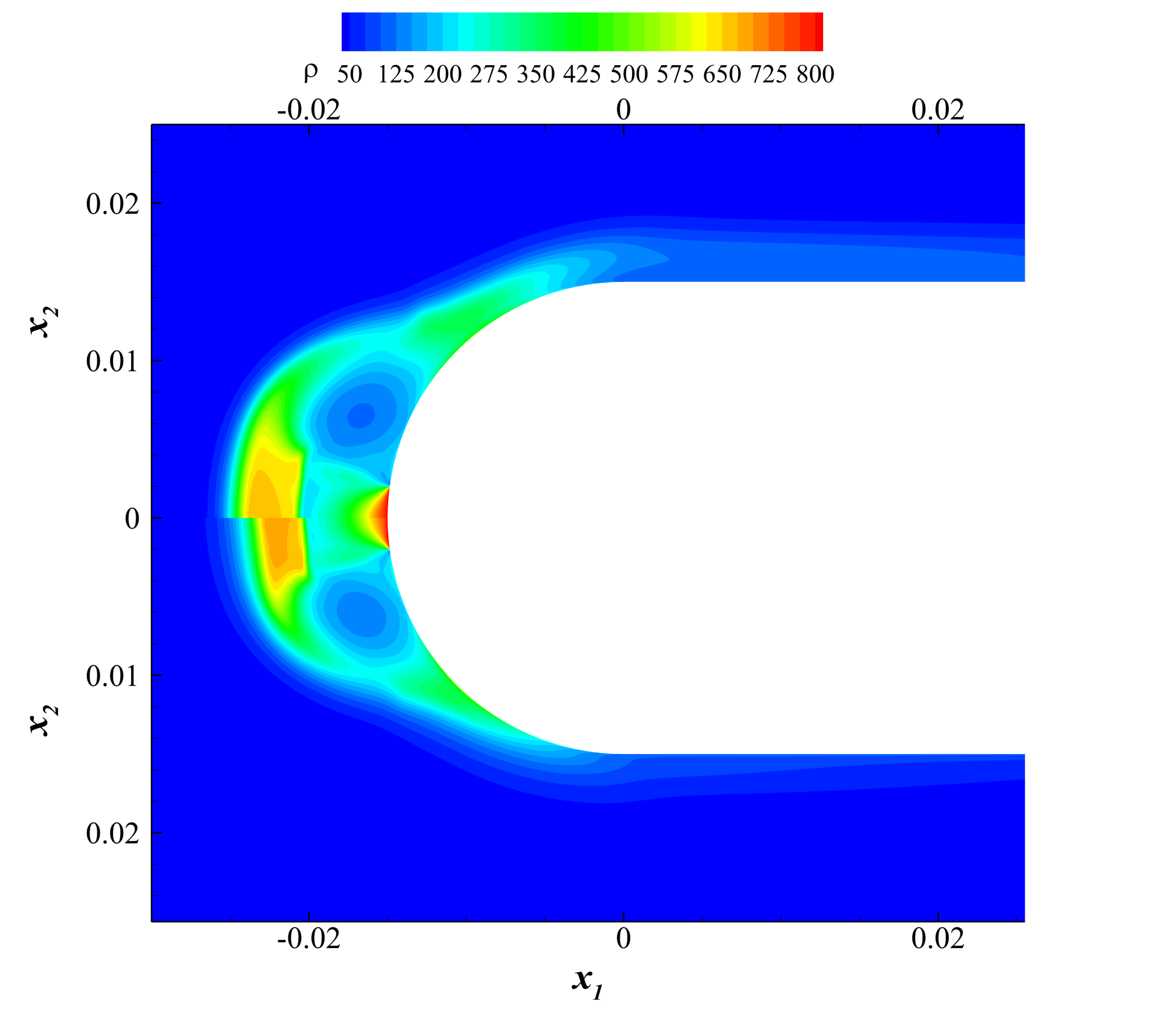}
     \caption{Density contour from GSIS (upper half) and GSIS-SST (lower half) at Kn=0.125. }
          \label{fig:JetHeatFluxResults}
\end{figure} 
 
When the turbulence is considered, large deviation between the GSIS and GSIS-SST could be spotted on the blunt part. At Kn=0.01, the GSIS-SST curve shows 70\% higher reattachment peak heat flux (compared to the GSIS one at same Kn) at around $x_1=0.01$ m. On the flat part of the model, GSIS-SST constantly predicts a 25$\sim$35\% higher $q$, no sign of deterioration of thermal protection range. The GSIS-SST curve from Kn=0.125 presents a 24\% higher reattachment peak heat flux and it greatly deviates from the GSIS one since $x_1=0~\text{m}$. The $q$ curve of GSIS-SST climbs almost linearly with $x_1$ on the flat part of the model, and resulting into a maximum relative deviation as high as {190\%}, implying an extremely fast degradation of the thermal protection effect. This could also be seen from the temperature (figure~\ref{fig:KnContourCP}) and density contour (figure~\ref{fig:JetHeatFluxResults}) at Kn=0.125, as the lower temperature and high density jet flow layer is much more limited in the GSIS-SST than GSIS. With more dispersed jet flow layer and hotter environment at Kn=0.125, the rest of the model (from $x_1=0$ m) in GSIS-SST is thus quickly heated.



\section{Conclusions}\label{sec:Conclusion}


A multiscale method, which is built upon the Boltzmann kinetic equation to describe the rarefied gas dynamics and the $k$-$\omega$ SST model to capture the turbulence effect, has been established to resolve gas flow problems from highly rarefied regime down to fully turbulent continuum regime. Asymptotic analysis and numerical simulations have proven that, when the Knudsen number is small (or the Reynolds is large), the GSIS-SST model recovers the standard SST model for turbulence modeling and reproduces the  boundary layer profiles in experiment, while when the Knudsen number is large, the GSIS-SST model recovers the Boltzmann kinetic equation. 
It should be noted that, while GSIS exactly solves the Boltzmann kinetic equation for laminar rarefied gas flows, the SST is an approximate model for turbulence viscosity. Therefore, the overall accuracy is affected by the SST model. Fortunately, the turbulence models are well studied, where abundant models/parameters can be optimized for specific problems.

With the proposed GSIS-SST model, the interaction between turbulent and rarefied gas flows has been analyzed. That is, for opposing jet problem under hypersonic rarefied flows, turbulence effect predicted by the SST model promotes the momentum and energy exchange between the jet flow and the free flow, considerably enhancing the jet diffusion and reducing the heat reduction range of the jet. 
For example, the surface heat flux predicted by the GSIS-SST may be approximately twice as high as that of the GSIS. The predicted turbulence also significantly increases reattachment surface pressure and heat flux, and such difference enlarges when the Kn number gets lower. 
These indicate the necessity of considering the turbulence effect of the jet even under rarefied free flow condition.

The significance of this study is twofold. Firstly, despite many researchers' unsuccessful attempts to find the rarefaction effects in turbulence, our research has presented, for the first time, numerical evidence of coexisting turbulent and rarefied gas flows with significant differences in experimental observables. These flows defy characterization by conventional turbulence models and cannot be accurately represented by laminar Boltzmann solutions.
Secondly, this study has provided a viable framework for advancing our understanding of the interaction between turbulent and rarefied gas flows, based on which more new flow phenomena can be explored. In addition to its application in hypersonic aerodynamics, the GSIS-SST may be applied in inertial confinement fusion to understand the interaction of turbulence in low-temperature high-density region and rarefied flows in high-temperature region with large mean free path of electron/ions~\citep{ICF2018}.

\section*{Acknowledgments} 
This work is supported by the National Natural Science Foundation of China (12172162). The authors thank Yanbing Zhang for his help with the GSIS code.

\section*{Declaration of interests} 
The authors report no conflict of interest.

\bibliographystyle{jfm}
\bibliography{Literature}

\end{document}